
\input harvmac.tex
\input epsf

\def\DefWarn#1{}
\noblackbox
\def\inbar{\,\vrule height1.5ex width.4pt depth0pt}
\def\IC{\relax\hbox{$\inbar\kern-.3em{\rm C}$}}
\def\IR{\relax{\rm I\kern-.18em R}}
\font\cmss=cmss10 \font\cmsss=cmss10 at 7pt
\def\IZ{\relax\ifmmode\mathchoice
{\hbox{\cmss Z\kern-.4em Z}}{\hbox{\cmss Z\kern-.4em Z}}
{\lower.9pt\hbox{\cmsss Z\kern-.4em Z}}
{\lower1.2pt\hbox{\cmsss Z\kern-.4em Z}}\else{\cmss Z\kern-.4em Z}\fi}
 \def\zb{{\bar{\vphantom\i z}}}

\def\CS{{\cal S}}\def\CP{{\cal P}}
\def\CM{{\cal M}}\def\CV{{\cal V}}
\def\p{\partial}

\def\pb{\overline{\partial}}

\def\lapl{\,\raise.5pt\hbox{$\mbox{.09}{.09}$}\,}

\font\manual=manfnt \def\dbend{\lower3.5pt\hbox{\manual\char127}}

\def\figin{\epsfcheck\figin}\def\figins{\epsfcheck\figins}
\def\epsfcheck{\ifx\epsfbox\UnDeFiNeD
\message{(NO epsf.tex, FIGURES WILL BE IGNORED)}
\gdef\figin##1{\vskip2in}\gdef\figins##1{\hskip.5in}
\else\message{(FIGURES WILL BE INCLUDED)}%
\gdef\figin##1{##1}\gdef\figins##1{##1}\fi}

\def\figinsert{\goodbreak\midinsert}
\def\ifig#1#2#3{\DefWarn#1\xdef#1{fig.~\the\figno}
\writedef{#1\leftbracket fig.\noexpand~\the\figno}%
\figinsert\figin{\centerline{#3}}\medskip\centerline{\vbox{\baselineskip12pt
\advance\hsize by -1truein\noindent\footnotefont{\bf Fig.~\the\figno:} #2}}
\bigskip\endinsert\global\advance\figno by1}

\catcode`\@=11   
\def\xeqn{\expandafter\xe@n}\def\xe@n(#1){#1}
\def\xeqna#1{\expandafter\xe@na#1}\def\xe@na\hbox#1{\xe@nap #1}
\def\xe@nap$(#1)${\hbox{$#1$}}
\def\eqns#1{(\e@ns #1{\hbox{}})}
\def\e@ns#1{\ifx\und@fined#1\message{eqnlabel \string#1 is undefined.}%
\xdef#1{(?.?)}\fi \edef\next{#1}\ifx\next\em@rk\def\next{}%
\else\ifx\next#1\xeqn#1\else\def\n@xt{#1}\ifx\n@xt\next#1\else\xeqna#1\fi
\fi\let\next=\e@ns\fi\next}
\catcode`\@=12

\def\inbar{\,\vrule height1.5ex width.4pt depth0pt}
\def\IB{\relax{\rm I\kern-.18em B}}
\def\IC{\relax\hbox{$\inbar\kern-.3em{\rm C}$}}
\def\ID{\relax{\rm I\kern-.18em D}}
\def\IE{\relax{\rm I\kern-.18em E}}
\def\IF{\relax{\rm I\kern-.18em F}}
\def\IG{\relax\hbox{$\inbar\kern-.3em{\rm G}$}}
\def\IH{\relax{\rm I\kern-.18em H}}
\def\II{\relax{\rm I\kern-.18em I}}
\def\IK{\relax{\rm I\kern-.18em K}}
\def\IL{\relax{\rm I\kern-.18em L}}
\def\IM{\relax{\rm I\kern-.18em M}}
\def\IN{\relax{\rm I\kern-.18em N}}
\def\IO{\relax\hbox{$\inbar\kern-.3em{\rm O}$}}
\def\IP{\relax{\rm I\kern-.18em P}}
\def\IQ{\relax\hbox{$\inbar\kern-.3em{\rm Q}$}}
\def\IR{\relax{\rm I\kern-.18em R}}
\font\cmss=cmss10 \font\cmsss=cmss10 at 7pt
\def\IZ{\relax\ifmmode\mathchoice
{\hbox{\cmss Z\kern-.4em Z}}{\hbox{\cmss Z\kern-.4em Z}}
{\lower.9pt\hbox{\cmsss Z\kern-.4em Z}}
{\lower1.2pt\hbox{\cmsss Z\kern-.4em Z}}\else{\cmss Z\kern-.4em Z}\fi}
\def\IGa{\relax\hbox{${\rm I}\kern-.18em\Gamma$}}
\def\IPi{\relax\hbox{${\rm I}\kern-.18em\Pi$}}
\def\ITh{\relax\hbox{$\inbar\kern-.3em\Theta$}}
\def\IOm{\relax\hbox{$\inbar\kern-3.00pt\Omega$}}

\def\CM {{\cal M}}
\def\CN {{\cal N}}
\def\CR {{\cal R}}

\def\CF {{\cal F}}
\def\CP {{\cal P }}
\def\CL {{\cal L}}
\def\CV {{\cal V}}
\def\CO {{\cal O}}

\def\CE {{\cal E}}
\def\p {\partial}
\def\CS {{\cal S}}

\def\pb{\bar{\partial}}

\def\bb {\bar{b}}
\def\cb {\bar{c}}
\def\zb {\bar{z}}
\def\wb {\bar{w}}
\def\log {{\rm log}}

\def\sdtimes{\mathbin{\hbox{\hskip2pt\vrule height 4.1pt depth -.3pt width
.25pt
\hskip-2pt$\times$}}}

\Title{\vbox{\baselineskip12pt\hbox{hep-th/9305139}\hbox{YCTP-P12-93}}}
{\vbox{\centerline{Finite in All Directions}}}
\bigskip

\centerline{Gregory Moore}
\bigskip\centerline{moore@castalia.physics.yale.edu}
\smallskip\centerline{Dept.\ of Physics}
\centerline{Yale University}
\centerline{New Haven, CT \ 06511}
\bigskip
\bigskip
\centerline{\it Dedicated to the memory of Feza G\"ursey}
\bigskip
\bigskip
\noindent
We study toroidal compactifications of string theories
which include compactification of a timelike coordinate.
Some new
features in the theory of toroidal compactifications
arise. Most notably, Narain moduli space does not exist
as a manifold since the action of duality on background
data is ergodic. For special compactifications certain
infinite dimensional symmetries, analogous to the
infinite dimensional symmetries of the $2D$ string are
unbroken. We investigate the consequences of these
symmetries and search for a universal symmetry which
contains all unbroken gauge groups.
We define a flat connection on the moduli space of
toroidally compactified theories. Parallel transport
by this connection leads to a formulation of broken
symmetry Ward identities. In an appendix this parallel
transport is related to a definition of conformal
perturbation theory.

\bigskip

\Date{May 24, 1993}

\noblackbox

\newsec{Introduction}

Symmetry has often been an important guide in finding
the fundamental formulation of physical theories.
Recent progress in closed string field theory (CSFT)
suggests that we are now much closer
to defining a theory which deserves the name {\it string theory}
\ref\bz{B. Zwiebach, ``Closed string field theory:
quantum action and the B-V master equation''
(hep-th/9206084), Nucl. Phys. B390 (1993) 33.}.
Unfortunately, the current formulation
is unwieldy and must be regarded as
an existence proof that covariant CSFT exists.
The present work is motivated by the hope that
a renewed investigation of the large symmetry algebras
appearing in string theory will lead to a better
understanding of CSFT. Common sense suggests that the
formulation of CSFT should drastically simplify around a
very symmetric background.

If $\CC$ is a CFT background of $c=26$ the ghost
number one BRST cohomology $H^1(\CC)$ may be
regarded as the Lie algebra of (inner) automorphisms of
the bosonic string background.
For example, in uncompactified
Minkowski space $\IR^{1,25}$ the
Lie algebra is $\IR^{26}\oplus \IR^{26}$ corresponding to
translations and dual translations. (The full Poincar\'e
symmetry does not arise as inner automorphisms.)
Recent investigations
of two-dimensional target spaces have revealed that
some $2D$ backgrounds have
{\it infinite-dimensional} unbroken symmetry algebras, e.g.,
algebras of area-preserving and volume-preserving
diffeomorphisms
\ref\witzwie{E. Witten and B. Zwiebach, ``Algebraic Structures and
Differential Geometry in 2D String Theory'' (hep-th/9201056),
Nucl. Phys. B377 (1992) 55.}.

An essential ingredient in the new constructions of
infinite-dimensional unbroken symmetries is the presence
of negative-dimension vertex operators
from the Liouville sector. In
general, the existence of large unbroken
symmetries is connected to the strange nature of
time in string theory.

In this
paper we will present further examples of backgrounds
with infinite-dimensional unbroken symmetries. Our technique
is elementary: we consider toroidal compactification of
all spacetime coordinates.
We are regarding compactified time as an
unphysical ground state, analogous to the ground state
$\langle \phi \rangle=0$ in electroweak theory.
Our philosophy is that underlying symmetries can
become manifest in backgrounds described by Higgs
vev's which differ from those chosen by nature.
Some of the unbroken symmetries we
encounter were proposed as being of fundamental
importance to string theory several years ago
\ref\gddrd{P. Goddard and D. Olive, ``Algebras, lattices
and strings,'' in {\it Vertex Operators in Mathematics and
Physics, Proceedings of a Conference} eds. J. Lepowsky,
S. Mandelstam, I.M. Singer, Springer}
\ref\witspec{E. Witten, ``Topological tools in ten
dimensional physics,'' in {\it Unified String Theories},
M. Green and D. Gross, eds. World Scientific}.

In outline, the main line of development in the
paper is the following:
We review toroidal compactification and discuss
some subtleties that arise upon compactification of
a timelike coordinate. We briefly describe the behavior
of the BRST cohomology and some enhanced symmetry points.
We show that there is a distinguished point in the
moduli space of toroidal compactifications which has, in
some sense, maximal symmetry.  We show how our
considerations naturally suggest a ``universal symmetry
algebra'' for toroidal compactifications.
In spontaneously broken
gauge theory Ward identities continue to hold even in
the broken phase. Therefore there must be
broken Ward identities even at points where the
enhanced symmetries are not present. In section nine
we propose a formulation of the broken Ward identities.
Along the way we are forced to clarify several points
about conformal perturbation theory and about the
nature of spontaneously broken symmetries in
toroidally compactified string theories.
We conclude with some far-fetched speculations.

This long paper is full of digressions, philosophical
remarks, speculations, proposed future directions etc.
For those in a hurry we provide a:

$\underline{Summary\ of\  new\ results}$

Section 2.5: New features of toroidal compactification of timelike
coordinates. Proposition 3 shows that duality does not
always act on sigma-model data. Proposition 5 shows that
the action of duality on the space of toroidally compactified
theories is ergodic.  These features are thoroughly studied
in an example in section 2.6.

Section 4.3: Proposition 8 shows that there are uniquely
distinguished string compactifications, depending only on the
number of gauged worldsheet supersymmetries. They are
distinguished by the requirement that the closed string
exactly factorize as a product of open strings.

Section 6: Proposition 10 constructs a candidate universal
symmetry for the enhanced symmetries in toroidal compactification.

Section 7.1: Defines globally a
parallel transport for the statespace of conformal field
theories over the moduli space of lattices.

Section 9.1: Propositions 16,17 formulate a set of
broken Ward identities for the enhanced symmetries
of toroidal compactifications.

Appendix A: Gives a complete contact term prescription to
make sense of conformal perturbation theory for relating
two toroidal compactifications. This involves a
preferred choice of coordinates on Narain moduli space
identified in Proposition 13.

We also include some tangential results which seem
interesting to us. These include:

Section 2.7: Proposition 7 gives an approximate fundamental
domain for the action of duality in Euclidean compactifications.

Section 4.4: Proposition 9 shows that compactifications
with Monster symmetry are not dense.

Section 8.2: Proposition 14 establishes a simple formula
for the cohomology class of string densities for
symmetry states.

\newsec{Toroidal Compactifications}

\subsec{Lattices}

Toroidal compactifications in $n+1$ dimensions are based on
lattices in $\IR^{n+1}\oplus \IR^{n+1}$ which are
even and unimodular ($=$ self-dual) with respect to
\eqn\dfdchk{\check D\equiv \pmatrix{0& 1\cr 1&0\cr}  .}
In order to consider the lattices as sets of points
in $\IR^{2n+2}$
we introduce a space of generator-matrices. Let
\eqn\metric{\eqalign{
 D=&\pmatrix{\eta &0\cr 0&-\eta\cr} \cr
\eta^{ab}_M=Diag\{-1,+1^{n}\}\qquad & \qquad
\eta^{ab}_E=Diag\{+1^{n+1}\} \cr}
}
in the case of Minkowskian and Euclidean compactifications,
respectively and consider
\eqn\defmht{
\CM = \{\CE\in GL(2n+2;\IR):  \CE^{tr}\cdot D \cdot \CE=\check D \}
}
$\CE$ defines a lattice with basis:
$(\CE^A_{\ I})$ where $A$
parametrize components and $I$ label basis
vectors.

Of course, $D$ and $\check D$ are similar:
\eqn\smtrm{\eqalign{
S D S = \check D &\qquad
S={1\over \sqrt{2}}\pmatrix{\eta & 1 \cr 1 & -\eta\cr}\cr
S&=S^{tr}=S^{-1}\cr}
}
When we need to distinguish Euclidean and Minkowskian cases
we use the notation $S_E$, $S_M$. Using \smtrm\
we may identify $\CM$ with orthogonal groups
\foot{We denote $O(Q;\IR)$ for the orthogonal
group determined by the quadratic form $Q$:
$$O(Q;\IR)=\{ g: g^{tr} Q g=Q\} $$  }
\eqn\orthgp{\eqalign{
 \CM \cdot S &= O(D;\IR)\cr
S \cdot \CM & = O(\check D;\IR)\cr}
}
Hence $\CM$ admits a left $O(D)$ and a right
$O(\check D)$ action.

A central theorem of the subject
states that the moduli space of even unimodular
lattices is (See
\ref\conway{J.H. Conway and N.J.A. Sloane, {\it Sphere Packings,
Lattices, and Groups} Springer-Verlag, 1993}\   ch. 15.)
\eqn\lattices{
\IL \equiv  \CM/O(\check D;\IZ)
\cong O(\check D;\IR)/O(\check D;\IZ)
}

\subsec{Abstract construction}

We construct a bundle of CFT statespaces
$\CH\to \IL$ as follows.

We form left and right Heisenberg-algebras with respect to
the quadratic
form $+\eta$ from the loop algebra $h=Map(S^1,\IR^{n+1})$.
Splitting $h=h^+\oplus h^0\oplus h^-$
as usual we have the state space
\eqn\statspi{
\CH_\Gamma=S(h_L^-)\otimes S(h_R^-)\otimes \IC[\Gamma]
}
where $S$ denotes the symmetric algebra, the last
factor is the group algebra of $\Gamma\in\IL$.

More concretely, consider the
dimension one currents:
\eqn\defyis{\eqalign{
 i \p Y^a(z)&=\sum \beta_n^a z^{-n-1}\cr
i \pb Y^a(z)&=\sum \bar\beta_n^a \bar z^{-n-1}\cr}
}
with
\eqn\defosc{\eqalign{
[\beta^a_n,\beta^b_m] &=\eta^{ab} n \delta_{n+m,0}\cr
[\bar\beta^a_n,\bar\beta^b_m] &=\eta^{ab} n \delta_{n+m,0}\cr}
}
If $\Gamma\in\CM$ we denote  $\Gamma_L=\pi_L(\Gamma)$,
$\Gamma_R=\pi_R(\Gamma)$, and vectors $(p_L;p_R)\in \Gamma$.
 In general $\Gamma_{L,R}$ are not lattices but
quasicrystals.
We denote the left and right Fock space built on
the momentum vacuum $|p_L\rangle, |p_R\rangle$
by $\CF_{p_L}$,$\bar\CF_{p_R}$.
The state $|p_L;p_R\rangle$ is created by $e^{i p_L\cdot Y(z)} e^{i p_R\cdot
Y(\bar{z})}C[p_L;p_R]$ where $C$ is a cocycle operator.
To any point $\CE\in \CM$ we associate the
CFT with statespace:
\eqn\statespace{\CH_\Gamma =\oplus_{(p_L;p_R)\in \Gamma}\CF_{p_L}
\otimes \bar\CF_{p_R} }
which is identical to \statspi.
Up to a rearrangement of terms in the direct
sum this only depends on $\Gamma=[\CE]\in \IL$ so we
can write $\CH_\Gamma$.

The above conformal field theories carry a natural action of
$O(\eta)\times O(\eta)$, left and right Lorentz transformations,
which change the lattice but preserve the correlation
functions. We wish to identify  CFT's related by
such Lorentz transformations to obtain the moduli space of CFT's.
More formally, if $\Gamma_1=g\cdot \Gamma_2$ are related by
$g\in O(\eta)\times O(\eta)$ there is a corresponding
isomorphism of vertex operator algebras
$U[g]:\CH_{\Gamma_1}\to \CH_{\Gamma_2}$ such that the
maps
\eqn\opfrmmaps{
\CP_{h,n}\to \CH_\Gamma^{\otimes n}
}
defined by the operator formalism (i.e., the Segal functor for
this CFT) satisfy:
\eqn\diagrami{
\matrix{ & &\CH_{\Gamma_1}\cr
&\nearrow& \cr
\CP_{h,n}& & \downarrow\cr
&\searrow& \cr
& & \CH_{\Gamma_2}\cr}
}
Here $\CP_{h,n}$ is the moduli space of genus $h$ surfaces with
$n$ coordinatized punctures.

With the equivalence \diagrami\ understood
the Narain moduli space of conformal field theories is
\eqn\narini{
\CN= \bigl(O(\eta)\times O(\eta)\bigr)\backslash
\CM/ O(\check{D};\IZ)
}

{\bf Remarks}:

1. If we also identify CFT's related by an exchange of
leftmovers for rightmovers (e.g. in the moduli space of
bosonic string theories) then the above holds with
$O(\eta)\times O(\eta)$ replaced by
$O(\eta)\times O(\eta)\sdtimes \IZ_2$.

2. If we work with lattices-with-basis defined by
$\CE\in \CM$ then the above identifications lead to the
moduli space:
\eqn\upfh{
\IH\equiv O(\eta)\times O(\eta)\backslash\CM\cong O(\eta)\times
O(\eta)\backslash O(D;\IR)
}
$\IH$ is the analog of the upper half plane in Riemann
surface theory.
The relation between theories from change of basis clearly
arises from the right-action of $O(\check D;\IZ)$ on $\IH$.
See the discussion of duality below.

3. We have not actually proved that $\CN$ is a true moduli
space since we have not shown there aren't other equivalences.
Locally this is obvious. Globally it is not so evident.

\subsec{Sigma-Model construction}

The previous discussion is a slightly unconventional
description of a very well-known and widely studied
class of theories -- the toroidally compactified theories.
We review this to set up some notation.

Let $X^\mu$ be coordinates on
the $n+1$-dimensional torus $X^\mu\sim X^\mu+2\sqrt{2} \pi$.
Discussions of
toroidal compactification usually begin with
the $\sigma$-model action
\eqn\sigmod{
S_{mink}={1\over 2 \pi} \int_0^{2 \pi} d \sigma \int d \tau
\p_- X^\mu E_{\mu \nu}\p_+ X^\nu
}
where $\p_\pm =\half (\p_{\tau}\pm \p_{\sigma})$ and the
worldsheet has Minkowskian signature.
The matrix $E$, referred to as the compactification data,
belongs to the space of $\sigma$-model backgrounds
\eqn\cmpctdta{
\CB=\{ E | E=G+B, signature(G)=\eta \}
}
where we always write the decomposition of $E$ into
symmetric and antisymmetric parts as $E=G+B$.
We assume $G$ is an invertible quadratic form
of signature $\eta$.

Canonical quantization leads in the standard way to
the oscillator expansions:
\eqn\oscexp{\eqalign{
\p_+ X^\nu &=\sum \bar\alpha_n^\nu e^{-i n(\tau+\sigma)}\cr
\p_- X^\nu &=\sum \alpha_n^\nu e^{-i n(\tau-\sigma)}\cr
[\alpha_n^\nu,\alpha_m^\mu]&=n \delta_{n+m,0} G^{\mu \nu}\cr
\alpha_0^\mu &= {1\over \sqrt{2}} G^{\mu \nu}(p_{\nu}+E_{\nu \rho} w^\rho)\cr
\bar\alpha_0^\mu &= {1\over \sqrt{2}} G^{\mu \nu}(p_{\nu}-E^{tr}_{\nu \rho}
w^\rho)\cr}
}
where $p_{\nu}, w^\rho$ are integers.

The Virasoro algebra is constructed as
$$L_n = \half \sum \alpha_{n-m}^\mu G_{\mu \nu} \alpha_m^\nu$$

The spectrum of the theory is therefore governed by the
operator
\eqn\hamil{
L_0+\bar{L}_0=\half \pmatrix{p&w\cr} \CR(E) \pmatrix{p\cr w\cr}
+N+\bar{N}
}
where
$$\CR(E)=
\pmatrix{G^{-1}& G^{-1}B\cr -B G^{-1} & G-B G^{-1}B\cr} $$
and therefore sigma models associated to
$E\in\CB$  with equivalent spectra are
classified by  matrices $\CR(E)$ equivalent under
conjugation by $O(\check D)$.

\subsec{Relation between the formulations}

We now relate the abstract to the $\sigma$-model
construction.
If $e^{~a}_\mu$ is a vielbein for
$G_{\mu \nu}$ then $\beta_n^a=e^{~a}_\mu \alpha_n^\mu(E)$
is canonically normalized as in \defosc.
In principle we could use different
vielbeins for left- and right-movers.
It follows from \oscexp\ that the lattice of zero-modes,
that is, the lattice of eigenvalues of $(\beta_0^a;\bar \beta_0^a)$
is obtained from a generator matrix
\eqn\smgnmtx{
\CE(e_1,e_2,E)\equiv {1\over\sqrt{2}}\pmatrix{ e_1^{a \mu}&
e_1^{a \mu} E_{\mu \nu}\cr
e_2^{a \mu}& - e_2^{a \mu} E^{tr}_{\mu \nu}\cr}
}
Here $(e_{1,2})^a_\mu$ are  vielbeins for $G_{\mu \nu}$, and
$e^{a \mu}=e^a_{~ \nu}G^{\nu \mu}$.
We will denote the space of matrices of the form \smgnmtx\
by $\CM_{\sigma}$.
A simple direct computation shows that $\CM_{\sigma}$
is contained within the space $\CM$ of \defmht.
Therefore, by projection
there is a well-defined map $\psi:\CB\to \IH$.

The relation between all the spaces we have discussed may be
summarized in the following diagram:
\eqn\diagramii{
\matrix{ &  & \CM_{\sigma} & \hookrightarrow & &\CM &  &\cr
  &\nearrow&    &   & \swarrow &  &\searrow & \cr
  \CB &  & {\buildrel \psi\over\longrightarrow }&\IH& & & &\IL\cr
         &        &        &   &\searrow& &\swarrow& \cr
         &         &     & & & \CN & &\cr}
}

This formulation makes the action of duality
on $\CB$ manifest.
The CFT only depends on the
choice of lattice, not of its basis, and lattices
are parametrized by $\IL=\CM/O(\check D;\IZ)$.
Therefore we study the {\it right} action of $O(\check D;\IZ)$
on $\IH$. We may attempt to find a right $O(\check D;\IZ)$
action on
compatification data $\CB$ by setting:
\eqn\duality{\eqalign{
A&=\pmatrix{a&b\cr c&d\cr}\in O(\check D;\IZ)\cr
\CE(e_1,e_2,E)\cdot A & =
\CE(\tilde e_1,\tilde e_2,\tilde E)\cr}
}
If $\tilde E$ exists such that the \duality\ is
satisfied then it is easily computed.
Using \smgnmtx\ and
equating the $11$ and $12$ blocks in \duality\
leads to the pair of equations:
\eqn\daulii{\eqalign{
e_1 G^{-1}(a + E c) & = \tilde e_1 \tilde{G}^{-1}\cr
e_1 G^{-1}(b + E d) & = \tilde e_1 \tilde{G}^{-1} \tilde E\cr}
}
Dividing the two equations we get:
\eqn\dltyact{
E\to \tilde{E}=(a+E c)^{-1}(b+E d) \qquad .}
{}From the abstract
formulation we see that we are simply making a
change of basis in $\CH_\Gamma$. Thus the Segal functor
is manifestly invariant, that is, all correlators are
duality covariant.

\noindent
{\bf Remark}: The full duality group was first identified in
\ref\givrab{A. Giveon, E. Rabinovici, and G. Veneziano, Nucl.
Phys. {\bf B322}(1989)167;  A. Giveon, N. Malkin, and E. Rabinovici,
Phys. Lett. {\bf 220B} (1989)}
\ref\shapere{A. Shapere and F. Wilzcek, Nucl. Phys.
{\bf B320}(1989)669}\
as $O(d,d;\IZ)$. In these
references emphasis is placed on
the matrix $\CR(E)$ appearing in \hamil.
The matrix $\CR(E)$ is related to $\CE$ by
\eqn\zerhami{
\CR(E)=\CE(e_1,e_2,E)^t\cdot
\pmatrix{\eta&0\cr 0& \eta\cr}\cdot \CE(e_1,e_2,E)\in O(\check D;\IR)}
If \duality\ holds then
$\CR(E)\to A^t \CR(E) A$. We find it easier to work with
the ``squareroot'' $\CE(e_1,e_2,E)$.

Of course \dltyact\
will only make sense if $(a+E c)$ is invertible.
In the case of Euclidean signature compactification this
is true since, as we will now show, $\psi$ is a
diffeomorphism and hence every matrix
$\CE\in\CM$ is of the form $\CE(e_1,e_2,E)$.

We first need a technical result which will be used several
times below.

\noindent
{\bf Proposition 1}: The Iwasawa decomposition
$$O(\check D;\IR) = {\bf K}\cdot {\bf A}\cdot {\bf N} $$
into compact, abelian and nilpotent subgroups is given by
\eqn\iwasawai{
g=\half\pmatrix{R_1+R_2&R_1-R_2\cr R_1-R_2&R_1+R_2\cr} \cdot\pmatrix{A&0\cr
0&A^{-1}\cr}\cdot \pmatrix{N&N B\cr 0&(N^{tr})^{-1}\cr}
}
where $R_i\in O(n+1)$, $A$ is diagonal with positive
entries, $N$ is upper triangular with $1$ on the diagonal,
and $B$ is an arbitrary $(n+1)\times (n+1)$ antisymmetric
real matrix.

\noindent
{\it Proof}: The embedding of the maximal compact subgroup
is obtained by conjugating $\pmatrix{R_1&0\cr 0& R_2\cr}\in O(D)$
by $S_E$. The form of the abelian and nilpotent
groups are dictated by the condition for upper block-triangular matrices
to lie in $O(\check D;\IR)$. The rest follows from
counting dimensions. $\spadesuit$

\noindent
{\bf Proposition 2}: In the case of Euclidean compactifications
$\psi:\CB\to \IH$ is a diffeomorphism.

\noindent
{\it Proof}:
We apply the Iwasawa decomposition above.
Using
$\CM=S\cdot O(\check D;\IR)$ it follows from Proposition 1 that
any matrix  $\CE\in \hat \CM$ can be written
uniquely  as
\eqn\iwasawa{
\CE=\pmatrix{R_1&0\cr 0&R_2\cr}\cdot S \cdot\pmatrix{A&0\cr
0&A^{-1}\cr}\pmatrix{N&N B\cr 0&(N^{tr})^{-1}\cr}
}
where $R_i$,$A$,$N$, and
and $B$ are as in Proposition 1.
Using the decomposition
\iwasawa\ we may define a map $\pi: \CM\to \CB$ by
associating to $\CE\in\CM$ the compactification data:
\eqn\efriwa{
\pi(\CE)=E=N^{-1} A^{-2} (N^t)^{-1} + B \qquad . }
One easily checks that
$\pi(\CE(e_1,e_2,E))=E$. Since the Iwasawa decomposition
describes a diffeomorphism of $O(\check D;\IR)$ with
$K A N$ it follows that
$\CB$ is diffeomorphic to $O(n+1)\times O(n+1)\backslash O(n+1,n+1;\IR)$.
$\spadesuit$

\subsec{New features of timelike compactification
}

There are two radically new features of toroidal compactification
of timelike coordinates.\foot{Conversations with G.D. Mostow
and G. Zuckerman were instrumental
in arriving at the results in this section.}
First,
in the Minkowskian case there is no analogous
Iwasawa decomposition for arbitrary $O(D;\IR)$ matrices.
This leads to

\noindent
{\bf Proposition 3}. The space of $\sigma$-model
data $\CB_{1,n}$ maps to a {\it proper} (open) subset of $\IH$.

\noindent
{\it Proof}: Use topology. In the Minkowskian case we have
\eqn\cmpctmnk{
\CB_{1,n}=\{ E | E=G+B, signature(G)=(-1,+1^n) \}
}
and $\CB$ is a vector bundle over the space of constant
Minkowskian metrics $GL(n+1,\IR)/O(1,n;\IR)$.
By a theorem of Mostow
\ref\mostow{G.D. Mostow, ``Covariant Fiberings of Klein
Spaces II,'' Amer. Jour. Math. {\bf 84} (1962) 466, and
refs. therein}\
we know that $GL(n+1,\IR)$ and $O(1,n;\IR)$ can be
compatibly decomposed as $K\cdot E$ where $K$ is a
maximal compact subgroup and $E$ is a Euclidean space,
so that $GL(n+1,\IR)/O(1,n;\IR)
\cong O(n+1)\times_{(O(1)\times O(n))}F$
where $F$ is an $n^2+n+1$ dimensional representation of
$O(1)\times O(n)$. Thus $\CB_{1,n}$ deformation retracts to
$\IR\IP^n$.

Now recall that
\eqn\deftorc{
\IH\equiv (O(\eta)\times O(\eta))\backslash \CM \qquad .}
Again applying Mostow's theorem we see that $\IH$
 is topologically
a vector bundle over $\IR\IP^{n}\times \IR\IP^{n}$.
Therefore $\CB_{1,n}$ is not diffeomorphic to $\IH$.
Since $\psi$ is locally $1-1$
it's image must be a proper open subset of $\IH$. $\spadesuit$

In fact, it is easy to give a criterion for when
$\CE\in\CM$ lies in $\CM_{\sigma}$:

\noindent
{\bf Proposition 4}.
$\CE\in\CM_{\sigma}$ iff $\CE_{11}$ is invertible, where
$\CE_{11}$ is defined by the block decomposition:
\eqn\ceblkfrm{
\CE=\pmatrix{\CE_{11}&\CE_{12}\cr \CE_{21}&\CE_{22}\cr}
}
In this case: $E=\CE_{11}^{-1} \CE_{12}$.

{\it Proof}: Writing the defining equations for
$\CE\in\CM$ gives several conditions on the blocks
$\CE_{ij}$. One then defines candidate
vielbeins from $e_1={1\over \sqrt{2}} \eta (\CE_{11}^{tr})^{-1}$,
$e_2={1\over \sqrt{2}} \eta (\CE_{21}^{tr})^{-1}$.
(The invertibility of $\CE_{21}$ follows from
that of $\CE_{11}$.)
The conditions that these are vielbeins for the same
metric, and that $\CE=\CE(e_1,e_2,E)$
 follows from the conditions on $\CE_{ij}$. $\spadesuit$

One striking consequence of Proposition 3 is that
the duality group $O(\check D;\IZ)$ does {\it not}
act on $\CB$. The reason is that when $\CE_{11}$ is
not invertible, the images of $\CE$ under $O(\check D;\IZ)$
will typically have $(\CE\cdot A)_{11}$ invertible.
This observation also answers in the affirmative
a question raised by Proposition 3, namely, whether
the projection $\CB\to \CN$ is actually onto.

{\bf Example}: Take
$$G=\pmatrix{0&1\cr 1&0\cr} \qquad B=\pmatrix{0&-1\cr 1&0\cr}\qquad
E=\pmatrix{0&0\cr 2&0\cr}$$
Then $E\to E^{-1}$ does not make sense. We study 2D
compactifications more thoroughly in the next section.

The second new feature of timelike compactification
is even stranger.
The action of
a noncompact semisimple Lie group on a space of the
form $G/\Gamma$ for $\Gamma$ an arithmetic discrete
group is typically {\it ergodic}. More precisely,
if $\Gamma\in G$ is a {\it lattice}, i.e., a discrete
subgroup such that
$\mu(G/\Gamma)<\infty$, where $\mu$ is the
Haar measure, then the action of a noncompact
subgroup $H\subset G$ is ergodic, i.e., the only
$H$-invariant subsets are of measure zero or total
measure. For reviews of this branch of mathematics
see, e.g.
\ref\ergodic{R.J. Zimmer, {\it Ergodic Theory and Semisimple
Groups}, Birkhauser}
\ref\margoulis{G. Margoulis, ``Dynamical and Ergodic
Properties of Subgroup Actions on Homogeneous Spaces with Applications to
Number Theory,'' in Proc. Int. Cong. Math. 1990, vol. 1 p.193}.
In the present
situation $O(\check D;\IZ)$ is a lattice in
$O(\check D;\IR)$. The left action of
$O(\eta)\times O(\eta)$, indeed, of any subgroup containing
boosts of arbitrarily large rapidity, on $\IL$ and the right-action
of the duality group $O(\check D;\IZ)$ on
$\IH$ are ergodic.
In fact, as a corrollary of some recent powerful theorems of
M. Ratner we may say much more:
\foot{This was explained to me by G.D. Mostow.}

\noindent
{\bf Proposition 5}:
Let $\CO_\Gamma$ be an
$O(\eta)\times O(\eta)$ orbit of $\Gamma\in \IL$, and let us
consider its closure $\bar{\CO}_\Gamma$ in $\IL$.
This set is either

\noindent
a.) a closed $O(\eta)\times O(\eta)$
orbit

or

\noindent
b.) $\IL$. That is, the orbit $\CO_\Gamma$ is dense in $\IL$.

Moreover, for almost every $\Gamma\in\IL$
possibility (b) is realized.

{\it Proof}:
Conjecture 2 of \margoulis\ proved as
Corollary B of Ratner's paper
\ref\ratner{M. Ratner, ``Raghunathan's Topological
Conjecture and Distributions of Unipotent Flows,''
Duke Math. Jour. {\bf 63}(1991)235}\
says that the closure of the
orbit of any unipotent group in $\IL$ is the orbit of some
connected group $H\subset O(\check D;\IR)$.
Moreover, this is true for any group generated
by unipotents.

In our case, $O(\eta)$ (and
hence $O(\eta)\times O(\eta)$) is generated
by unipotent flows. For example, for $O(\eta)$
we can choose lightcone directions $\pm$, and transverse
directions $j=1,\dots n-2$.
The Lorentz generators $M_{\pm j}$
have zero eigenvalues under the adjoint action and
hence exponentiate to give unipotent flows. On the
other hand, their commutators generate the full
Lorentz group.

Applying Ratner's theorem we conclude that
$\bar{\CO}_\Gamma$ is the orbit of some closed
subgroup $O(\eta)\times O(\eta)\subset H\subset O(\check D;\IR)$.
Again, by studying the adjoint representation we see that
if there are any ``off-diagonal'' generators of $H$ not in
$O(\eta)\times O(\eta)$, then commutators with
$O(\eta)\times O(\eta)$ will generate the whole group.
Thus $H=O(\eta)\times O(\eta)$ or $H=O(\check D;\IR)$,
yielding possibilities $a$ and $b$ above.
That possibility $(b)$ is almost always realized follows
from Theorem 11.1 of
\ref\mostowi{G.D. Mostow, Publ. I.H.E.S. {\bf 34}(1968) 96}.
$\spadesuit$

{\bf Remarks}:

1. We will
exhibit the ergodic action quite explicitly for the case of
compactification of $1+1$ dimensions in the next section.

2. The statements in Proposition 5 for the orbit
structure have analogs for the action of the arithmetic
group $O(\check D;\IZ)$ on $\IH$.

3. It follows from the above that the space $\CN$ does
not exist as a reasonable space. This result contradicts
some folklore. Several
developments might restore harmony: (1) It might
be that $\CN$ is embedded in some larger
space of theories, and it is just a pathological
subset. (2) It might be that the theories under
discussion are pathological, and should not really be
considered as CFT's. (See the discussion of
loop amplitudes in section 8.4.) Perhaps when time enters
nontrivially the moduli space of string theories is
typically smooth but closed timelike loops are a
source of pathology. (3) It might be that
we have uncovered an important aspect of the
space of time-dependent string backgrounds, and
that we will be forced to think about
noncommutative geometry on $\CN$. Time will tell.

\subsec{Example: Compactification of $1+1$ Dimensions}

It is quite well-known in Euclidean compactification that
one can be very explicit about two-dimensional targets
by use of the isomorphism of Lie algebras
$so(2,2)\cong sl(2)\times sl(2)$
\ref\dvv{R. Dijkgraaf, E. Verlinde, and H. Verlinde, ``On the
moduli spaces of conformal field theories with
$c\geq 1$,'' in {\it Perspectives in String Theory},
P. DiVechia and J.L. Petersen, eds.}
\shapere. In this
subsection we apply the isomorphism to our case.
We work out both Euclidean and Minkowskian cases because
the comparison is interesting.

$\underline{Group\ isomorphism.}$

Set
$$\eqalign{
\delta_1=\pmatrix{1&0\cr 0&0\cr} \qquad & \delta_2=\pmatrix{0&1\cr 0&0\cr}\cr
\delta_3=\pmatrix{0&0\cr 0&1\cr} \qquad & \delta_4=\pmatrix{0&0\cr -1&0\cr}\cr}
$$
so that
$$2 \det(x^\mu \delta_{\mu})=x^\mu \check{D}_{\mu \nu} x^\nu$$
Therefore we may define a homomorphism
\eqn\defhomo{\eqalign{
\psi:SL(2,\IR)\times SL(2,\IR) & \to O(D;\IR)\cr
\psi(A,B)^\mu_{\ \nu} &=\half \tr\bigl (A \delta_{\nu} B^{tr}
\delta_{\mu}^{tr}\bigr)\cr}
}
The group $O(\check D;\IR)$ has four components
and may be written as a disjoint union:
\eqn\disjunion{\eqalign{
O(\check D;\IR) &=O_0(\check D;\IR)\amalg
\bigl(O_0(\check D;\IR)\cdot P_L\bigr)
\amalg \bigl(O_0(\check D;\IR)\cdot T_L\bigr)
\amalg \bigl(O_0(\check D;\IR)\cdot P_L T_L\bigr)\cr
T_L&=S_M Diag\{-1,1,1,1\} S_M
=\pmatrix{0&0&1&0\cr 0&1&0&0\cr 1&0&0&0\cr 0&0&0&1\cr} \cr
P_L&=S_M Diag\{1,-1,1,1\} S_M=\pmatrix{1&0&0&0\cr 0&0&0&-1\cr 0&0&1&0\cr
0&-1&0&0\cr}\cr}
}

The kernel of $\psi$ is $\IZ_2$, generated by $(-1,-1)$, and the image is
the connected component of the identity: $O_0(\check D;\IR)$, so it
defines an isomorphism:
$${SL(2,\IR)\times SL(2,\IR)\over \IZ_2}\cong O_0(\check D;\IR)$$

A very useful property of the homomorphism $\psi$ is that
it is compatible with Iwasawa decompositions. That is,
decomposing $SL(2,\IR)$ as
$$SL(2,\IR)= K A N$$
where $K=SO(2)$, $A$ is diagonal, and $N$ is unipotent and
upper triangular, the images give the KAN subspaces in
$O(\check D;\IR)$ described in Proposition 1.

$\underline{Duality\ group.}$

Since
$P_L,T_L\in O(\check D;\IZ)$, with $P_L^2=T_L^2=1$, it follows
that $O_0(\check D;\IZ)$ is an index $4$ subgroup of $O(\check D;\IZ)$.
Using an explicit set of generators
\ref\givgen{A. Giveon, N. Malkin, E. Rabinovici,
Phys.Lett.B238:57,1990.} one may
examine the inverse images to conclude that
$\psi$ defines an isomorphism
\eqn\dulim{
{SL(2,\IZ)\times SL(2,\IZ)\over \IZ_2}\cong O_0(D;\IZ)
}
The full duality group is then given by
\eqn\exctseq{
1\to \IZ_2\times\IZ_2 \to O(\check D;\IZ) \to
{SL(2,\IZ)\times SL(2,\IZ)\over \IZ_2} \to 1
}
The group extension is defined by the action of $T_L,P_L$ as
outer automorphisms:
\eqn\outaut{\eqalign{
T_L: (A,A')& \to (J A' J, J A J)\cr
P_L: (A,A')& \to (A',A)\cr
J&=\pmatrix{0&1\cr 1&0\cr} \cr}
}

$\underline{\IH : Euclidean\ Compactification}$

The distinction between Euclidean and Minkowskian
compactifications first enters at this point.
We identify the space of Euclidean generator
matrices by
$$\CM=S_E\cdot O(\check D;\IR)$$
$S_E$ conjugates $O(2)\times O(2)$ to the maximal
compact subgroup of $O(\check D;\IR)$. Moreover
$K\backslash O(\check D;\IR)\cong
K_0\backslash O_0(\check D;\IR)$.
By the compatibility of Iwasawa decompositions we find that
\eqn\ismehh{
\IH\equiv (O(2)\times O(2))\backslash \CM\cong
(SO(2)\backslash SL(2,\IR))\times (SO(2)\backslash SL(2,\IR))
}
may be identified with the product of upper half planes.

$\underline{\IH : Minkowskian\ Compactification}$

Now we have
$$\CM=S_M\cdot O(\check D;\IR)$$
The left and right Lorentz groups $O(\eta)\times O(\eta)$
do {\it not} embed into the diagonal subgroup $A$,
but rather correspond (in $SL(2,\IR)\times SL(2,\IR)$)
to the conjugate group
$$\eqalign{
SO(1,1)\times SO(1,1) &\cong s A s^{-1}\times s^{-1} A s\cr
s &={1\over \sqrt{2}} \pmatrix{1&-1\cr 1& 1\cr}\cr}
$$
In this way we identify
\eqn\ehhmink{
\IH\cong ((\IZ_2\times \IZ_2\times \IZ_2))\sdtimes SO(1,1)\times
SO(1,1))\backslash SL(2,\IR)\times SL(2,\IR)
}
where the last $\IZ_2$ is $(-1,-1)$ and the first two are generated by
\eqn\morezts{
(\pmatrix{0&1\cr -1&0\cr},\pmatrix{0&-1\cr 1&0\cr})\qquad
(\pmatrix{0&-1\cr 1&0\cr},\pmatrix{0&-1\cr 1&0\cr})
}
These are present because, of the 16 components of the
left and right Lorentz groups only 4 are killed by passing
to the connected component.

$\underline{Interpretation\ of\ \IH_{Mink}}$

It will be useful below to have a clear geometrical interpretation
of $\IH_{Mink}$.

Conformal equivalence classes of flat Lorentzian tori in
$1+1$ dimensions may be specified by constant metrics,
which may in turn be obtained from the standard metric
in $1+1$ dimensions by compactifying on a lattice
with basis $e^1,e^2$.
Letting $e$ denote the matrix of zweibeins we
obtain the  Lorentzian metric
$$G= e^{tr}\pmatrix{-1&0\cr 0&1\cr} e$$
on the torus $(x {\rm mod } 1,t {\rm mod} 1)$.
Consequently ``Teichm\"uller space'' for flat Lorentzian signature tori
may be regarded as the homogeneous space
$\IZ_2\times SO(1,1)\backslash SL(2,\IR)$.
Consider a different Iwasawa decomposition:
\eqn\newiwa{
g=\pmatrix{\cosh \beta&\sinh \beta\cr  \sinh \beta&\cosh \beta\cr}
\pmatrix{1+n/2&-n/2\cr n/2& 1-n/2\cr}
\pmatrix{\cos \omega&\sin \omega\cr -\sin \omega&\cos \omega\cr}
}
where $n\in \IR$. Using \newiwa\
we identify the homogenous space
$\IZ_2\times SO(1,1)\backslash SL(2,\IR)$
as $\IR\times S^1$.
Thus, $\IH_{mink}$ is a product of cylinders with a
$\IZ_2$ identification. $\IH_{Mink}$ can
be  interpreted as the
Teichm\"uller space of two flat Lorentzian tori, which are
separately unoriented, but carry a relative
orientation.

We can relate this discussion to the standard discussion
of Teichm\"uller space for tori by introducing
two real ``modular parameters'':
\eqn\modprmtrs{\eqalign{
\tau_+ &= -\biggl({G_{12}\over G_{11}} + {\sqrt{|\det G|}\over G_{11}}\biggr)
={e_2^{~2}-e_1^{~2}\over e_1^{~1}-e_2^{~1}}=e^{-1}\cdot 1\cr
\tau_- &= -\biggl({G_{12}\over G_{11}} - {\sqrt{|\det G|}\over
G_{11}}\biggr)=-{e_2^{~2}+e_1^{~2}\over e_1^{~1}+e_2^{~1}}
=e^{-1}\cdot (-1)\cr}
}
In the last equalities we regard $e\in SL(2,\IR)$ acting
on the real line by M\"obius transformations.
It is manifest from this formula that global diffeomorphisms
act on $\tau_\pm$ by integral M\"obius transformations.

The modular parameters are not always good coordinates on
Teichm\"uller space. For example, one of $\tau_\pm$ will be infinite
when one of the basis vectors becomes a null vector. We
may remedy this problem by compactifying the
real line to $\bar{\IR}=\IR\cup \infty$. A second problem
is that the diagonal $\tau_+=\tau_-$ leads to degenerate
tori where both basis vectors are parallel null vectors.
By adding such degenerate tori we can make the
Teichm\"uller space itself a torus $S^1\times S^1$.
Excising the circle $\tau_+=\tau_-$ of degenerate
tori we restore the cylinder $\IR\times S^1$.

$\underline{Relation\ to\ \CB:\ Euclidean\ Case}$

We now construct the compactification
data associated to
an element of $SL(2,\IR)\times SL(2,\IR)$ by setting
$$\CE = S_E \psi(A,A') $$
and then calculating $E=\CE_{11}^{-1} \CE_{12}$ in
accord with Proposition 4.
We find:
\eqn\efrmsl{\eqalign{
E&= G+B\cr
B&={a b+cd \over a^2+c^2}\pmatrix{0 & -1\cr 1&0\cr}\cr
G&={1\over a^2+c^2}
\pmatrix{(b')^2+(d')^2 & -a' b'-c' d'\cr -a' b'-c' d'& (a')^2+(c')^2\cr} \cr}
}
{}From which we may form:
\eqn\rhotau{\eqalign{
\rho&=B_{12}+i\sqrt{\det G}= {d i -b\over -c i + a}=A^{-1}\cdot i\cr
\tau&=
{G_{12}\over G_{22}} +i {\sqrt{\det G}\over G_{22}}={d' i -b'\over -c' i +a'}
=(A')^{-1}\cdot i\cr}
}
which parametrize the upper half planes.

$\underline{Relation\ to\ \CB:\ Minkowskian\ Case}$

In an entirely similar way we set
$$\CE = S_M \psi(A,A') $$
and calculate $E$.
We find
\eqn\efrmsli{\eqalign{
E&= G+B\cr
B&={a b-cd \over a^2-c^2}\pmatrix{0 & -1\cr 1&0\cr}\cr
G&=-{1\over a^2-c^2}
\pmatrix{-(b')^2+(d')^2 & a' b'-c' d'\cr a' b'-c' d'& -(a')^2+(c')^2\cr} \cr}
}

We may now explicitly verify Proposition 3 since the formulae
in \efrmsl\ only make sense when $a^2-c^2\not=0$. Indeed,
\eqn\compeoo{
\det \CE_{11}=-\half (c^2-a^2)
}
for the component of the identity, and similarly for
other components.
Excising the submanifolds $a^2-c^2=0$. we see that the
projection of $\CM_{\sigma}$ to $\IH$ has four components,
depending on the signs of $a\pm c$.

Forming modular parameters as before we now have
\eqn\efrmslii{\eqalign{
\rho_\mp &\equiv B_{21}\mp \sqrt{|\det G|} \cr
\tau_\mp &\equiv - {G_{12}\over G_{22}}\mp {\sqrt{|\det G|}\over G_{22} }\cr}
}
In terms of the $SL(2,\IR)$ parameters we have:
\eqn\sltps{\eqalign{
\rho_- &= -{b+d\over c+a}=A^{-1}(-1)\cr
\rho_+ &={d-b\over a-c}=A^{-1}(+1)\cr
\tau_- &=(A')^{-1}(-1) \cr
\tau_+ &=(A')^{-1}(+1) \cr}
}
for $a^2>c^2$. For $a^2<c^2$ we exhchange
$\rho_-\leftrightarrow \rho_+$ and
$\tau_-\leftrightarrow \tau_+$.

These are modular parameters for a pair of flat spacetime
Lorentzian tori, in accord with our interpretation of
$\IH_{Mink}$. The tori which are missed in $\CM-\CM_{\sigma}$
are those for which the {\it dual} torus, parametrized
by $\rho_\pm$ has a null basis vector.

$\underline{Action\ of\ duality: Euclidean\ case}$

{}From the expression in \rhotau\ in terms of Mobius
transforms of $i$ we see that the rightaction of
the $SL(2,\IZ)\times SL(2,\IZ)$ subgroup of the
duality group is manifestly the Mobius action on the upper
half planes. Using \outaut\ we see that the
remaining generators act by
$T_L:(\rho,\tau)\to (1/\bar{\tau},1/\bar{\rho})$ and
$P_L:(\rho,\tau)\to (\tau,\rho)$. If $T_L$ is
composed with other duality generators it becomes
simply a spacetime parity transformation.

$\underline{Action\ of\ duality: Minkowskian\ case}$

Similarly,
using the representation \efrmslii\ we see that
a duality transformation
$(A,A')\in SL(2,\IZ)\times SL(2,\IZ)$ acts
on $(\rho_\pm,\tau_\pm)$ by Mobius transformations:
\eqn\mobact{
(\rho_\pm,\tau_\pm)\to (A^{-1}\cdot \rho_\pm,(A')^{-1}\cdot \tau_\pm)
}
as long as the image lies in the sigma-model set.
As discussed above, we should compactify the the
$\rho_\pm,\tau_\pm$ real lines and also admit
degenerate tori.
We explicitly verify that $\CB\to \CN$ is onto since
one can always transform infinity to a finite point.

The generators of the two remaining $\IZ_2$'s act as follows.
The transform of $(\rho_-,\rho_+,\tau_-,\tau_+)$ is
\eqn\dauliii{\eqalign{
T_L: (1/\tau_-,1/\tau_+,1/\rho_-,1/\rho_+)&\qquad {\rm if}\quad
a^2>c^2,(d')^2>(b')^2\cr
P_L: (\tau_-,\tau_+,\rho_-,\rho_+)&\qquad {\rm if}\quad
a^2>c^2,(a')^2>(c')^2\cr}
}
The other cases are similar.

$\underline{Ergodic\  duality}$

We can now show that duality acts ergodically by following
an analysis of E. Artin
\ref\artin{E. Artin, ``Ein mechanisches System mit quasiergodischen
Bahnen,'' Collected works, p. 499}, who showed that the
generic geodesic on the modular curve
$SO(2)\backslash SL(2,\IR)/SL(2,\IZ)$ is dense.
\foot{See the geometrical interpretation below.}

It suffices to consider the action of
$PGL(2,\IZ)$ on $(\rho_-,\rho_+)$.
WLOG, take $(\rho_-,\rho_+)$ to
lie in the domain $1<\rho_+, -1<\rho_-<0$.
For $\rho_\pm$ in this domain consider
the continued fraction expansions:
\eqn\cntdfrac{\eqalign{
\rho_+&=\langle a_0,a_1,a_2,...\rangle \equiv a_0 +
{1\over a_1 + {1\over a_2 + \cdots}}\cr
-\rho_-&=\langle 0,a_{-1},a_{-2},...\rangle \equiv
{1\over a_{-1} + {1\over a_{-2} + \cdots}}\cr}
}
Together these expansions define a
bi-infinite series of positive
integers
$$\CS=\cdots a_{-2}, a_{-1},a_0,a_1,a_2,\dots\qquad . $$
If we ``cut'' the series in two at some
point $a_n$ we can define two new continued fractions. Set
\eqn\cntdfr{\eqalign{
\rho_+^{(n)}&=\langle a_n,a_{n+1},a_{n+2},...\rangle \cr
-\rho_-^{(n)}&=\langle 0,a_{n-1},a_{n-2},...\rangle \cr}
}

{\bf Theorem}(Artin \artin).

a.) For all $n\in\IZ$ the pair $(\rho_-^{(n)},\rho_+^{(n)})$ are
$PGL(2,\IZ)$ transforms of $(\rho_-,\rho_+)$.

b.) All $PGL(2,\IZ)$ transforms which lie in the domain
$1<\rho_+, -1<\rho_-<0$ are given by \cntdfr.

Statement (a) is a straightforward consequence of
the usual recursion relations for continued fractions
\ref\keng{See, e.g., H.L. Keng, {\it Introduction to Number Theory},
Springer}.

Therefore, for almost all compactification data, the
$SL(2,\IZ)$ orbit will be dense. Indeed, if we
wish to approximate any pair
$$(\tilde \rho_+,\tilde \rho_-)=(\langle c_0,c_1,c_2,\dots\rangle,
\langle b_1,b_2,\dots\rangle)$$
we can look deep in the sequence
$\CS$. For generic sequences, and for any $m$, we can
find $n$ large enough so that $a_{n-m},\dots,a_{n+m}$ corresponds to
$b_m,b_{m-1},\dots b_1,c_0,c_1,\dots c_{m-1},c_m$. In this case the
image $(\rho_+^{(n)},\rho_-^{(n)})$ approximates
$(\tilde \rho_+,\tilde \rho_-)$ and the approximation can be
made arbitrarily good by increasing $m$.

If $\CS$ is not generic then it will have repeating
subsequences if we look deep enough.
In particular, if $a_j$ eventually becomes
zero for $|j|$ large and $\rho_\pm$ are rational then
the orbit is not dense. Correspondingly, the
$O(1,1)\times O(1,1)$ orbit in $\IL$ is closed.
This will be important later.

$\underline{Geometrical\ interpretation}$

There is a lovely geometrical picture for the
ergodic action we are describing.
\foot{See
\ref\sullivan{D. Sullivan, ``Discrete Conformal Groups
and Measurable Dynamics,'' B.A.M.S. {\bf 6}(1982)57}
for a recent review.}
The double  coset
$SO(2)\backslash SL(2,\IR)/SL(2,\IZ)$
is a Riemann surface $\Sigma$. It is a punctured sphere with
three orbifold points. We may regard $SL(2,\IR)/SL(2,\IZ)$
as the unit tangent bundle $T_1\Sigma$ in the natural
constant curvature metric. The left action of the
subgroup
$$\pmatrix{\cosh t& \sinh t\cr \sinh t& \cosh t\cr} $$
on $SL(2,\IR)/SL(2,\IZ)$ induces geodesic flow on $T_1\Sigma$.
The projection of the geodesics on $\Sigma$ are
obtained from the upper half plane $SO(2)\backslash SL(2,\IR)$
by projecting semicircles perpendicular to the real
axis, as in
\ifig\geodesic{A geodesic on the modular curve is obtained by
projecting a semicircle from the upper half plane. The geodesic
is specified by two real numbers $\rho_\pm$ up to simultaneous
$SL(2,\IZ)$ transformation.}{\epsfxsize 3in\epsfbox{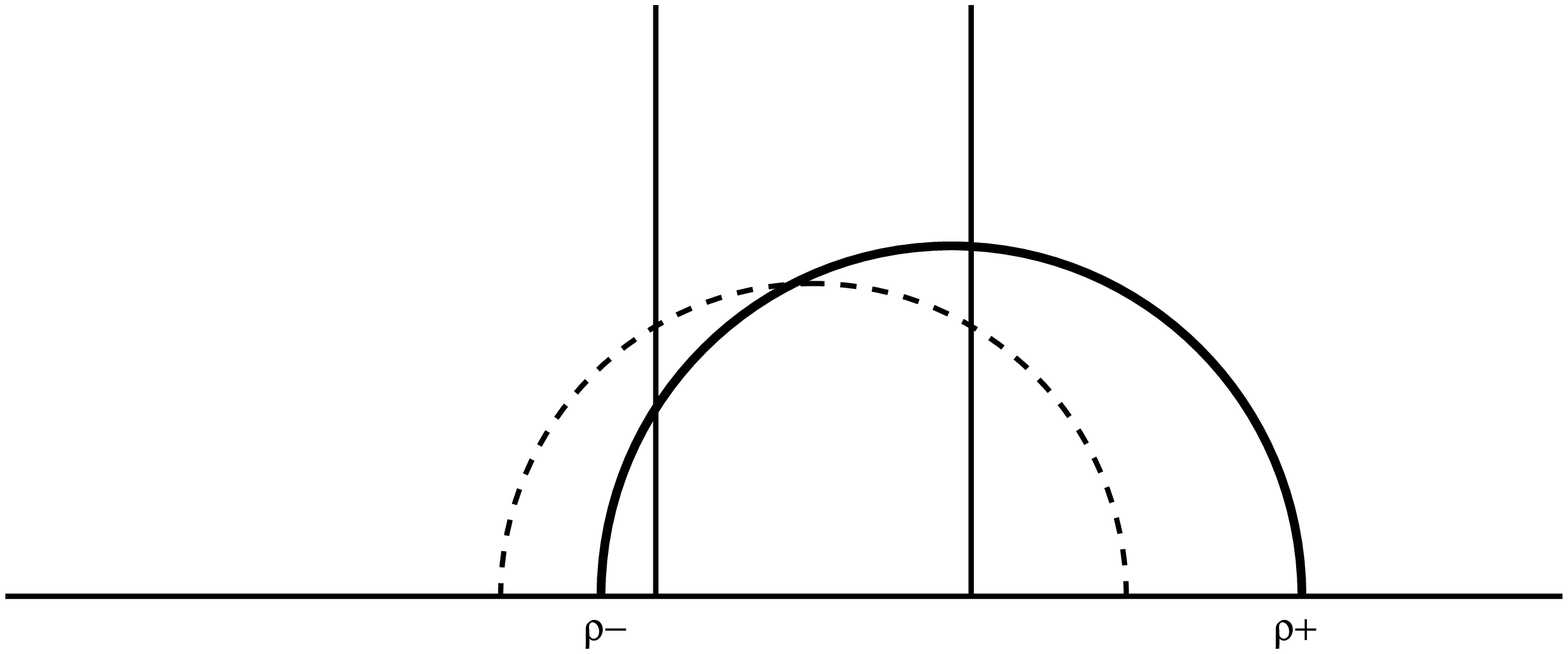}}

To obtain the complete geodesic we must look at all
$SL(2,\IZ)$ transforms of the semicircle.
The connection
between geodesics and a pair of real numbers, $\rho_\pm$ is
established by interpreting $\rho_\pm$ as the basepoints of a
semicircle in the upper half plane.  The significance of the
domain $1<\rho_+, -1<\rho_-<0$ is that for these circles there
is a nonzero intersection with the standard fundamental domain.
Artin's analysis
shows that almost all geodesics are dense and
characterizes those which are not. Note, for example, that
the geodesics with rational basepoints are closed non-dense
orbits. Other geodesics whose basepoints correspond to special
continued fraction expansions will also fail to be dense.

\subsec{Decompactifying}

Another major difference in the Euclidean and
Minkowskian cases is how one describes decompactification.

Although $\IL$ has finite measure it is a noncompact.
Indeed, the boundary components at infinity describe the
ways in which a Euclidean spacetime torus can be decompactified.
There exists a beautiful mathematical theory which describes
the structure of arithmetic quotients, such as  $\IL$, at
infinity. In this section we recall some of that theory
and apply it to our case.
\foot{I am indebted to Howard Garland for
explaining compactification of arithmetic quotients
to me.}
A useful reference for this subject is
\ref\borel{A. Borel, {\it Introduction aux groupes arithmetiques},
Hermann, Paris, 1969.}.

The main idea is easily explained using the arithmetic
quotient $T_1\Sigma = SL(2,\IR)/SL(2,\IZ)$ introduced above.
Introduce an Iwasawa decomposition:
$$g=kan=k \pmatrix{\lambda&0\cr 0& \lambda^{-1}\cr} \pmatrix{1&-x\cr 0&1\cr}$$
so that, with the right action on the upper half-plane
$SO(2)\backslash SL(2,\IR)$ we identify
\eqn\uhpl{
z=g^{-1}i= x+i \lambda^{-2}. }
Note that although the fundamental domain for the action
of $SL(2,\IZ)$ on $SO(2)\backslash SL(2,\IR)$ is relatively
complicated, we can easily describe an approximate
fundamental domain
in terms of ``Siegel sets.''
Let: $A_t= \{a\in A: \lambda^2<t \}$ and $N_t=\{n\in N: |x|<t\}$.
Then $\Sigma_{t,\half}=K\cdot A_t \cdot N_\half$
defines an approximate fundamental
domain for $t\geq {\sqrt{3}\over 2}$.
More formally, $SL(2,\IR)=\Sigma_{t,\half}\cdot \Gamma$,
where $\Gamma= SL(2,\IZ)$.

We can define what it means to go to infinity using the
Siegel sets. Clearly, from \uhpl\ we want
to let $\lambda^2\to 0$. Thus, a neighborhood of
infinity is defined by $\Sigma_{t,\half}$ for
$t$ small.

The generalization of this idea to
$\IL=O(\check D;\IR)/O(\check D;\IZ)$ is the following.
Returning to the Iwasawa decomposition of Proposition 1
we define a root system using the characters of the maximal
abelian
subgroup. If $A=Diag\{\lambda_1,\dots \lambda_n\}$ then a
set of simple roots is given by
\eqn\splrts{\eqalign{
\alpha_i &=\lambda_i/\lambda_{i+1} \qquad 1\leq 1\leq n-1\cr
\alpha_n &=\lambda_{n-1} \lambda_n\cr}
}
We first define an approximate fundamental domain.
The Siegel sets are defined by:
$\Sigma_{t,\omega}={\bf K}{\bf A}_t {\bf N}_{\omega}$ where
\eqn\siegseti{
{\bf A}_t=\{a\in {\bf A}: \alpha_i(a)<t, \qquad i=1,\dots n\}
}
and ${\bf N}_{\omega}$ is a compact fundamental
region for ${\bf N}/{\bf N}\cap \Gamma$, where $\Gamma$ is
the arithmetic group ($=O(\check D;\IZ)$ in our
example). For example, we can, and will,
define
$${\bf N}_{\omega}=\{ g=\pmatrix{N&NB\cr 0&(N^{tr})^{-1}\cr}:
|n_{ij}|\leq \half, |B_{ij}|\leq \half \}\qquad . $$
where $n_{ij}$ denote the matrix elements of $N$ above the
diagonal.

\noindent
{\bf Proposition 6}: If $t\geq {\sqrt{3}\over 2}$ then
for $G=O(\check D;\IR)$
\eqn\siegsetii{
G={\bf K}{\bf A}_t {\bf N}_{\omega}\cdot \Gamma
}

\noindent
{\it Proof}: This is a straightforward
generalization of the discussion on
pp. 13-15 of \borel. We refer to \borel\
for notation. One defines $\Phi(g)$ in the
same way with respect to the standard
Euclidean norm on $\IR^{2n}$. ${\bf K}$
is orthogonal with respect to this norm, and
Lemma 1.5 of \borel\ goes through unaltered. Likewise
the inductive proof of 1.6 of \borel\ needs only two
small alterations. The first part
of the inductive step may be established
by the explicit description of two-dimensional
compactification of the previous subsection.
Second, to embed $O(\check D_{n-1},\IR)$
into $O(\check D_{n},\IR)$ we conjugate with
the permutation taking
$$x_2\to x_3\to \cdots \to x_n\to x_{n+1}\to x_2$$
and use the obvious block diagonal embedding.
$\spadesuit$

We may apply this to Euclidean compactifications using
\efriwa\ to show that there is a simultaneous lower bound
on all the ``radii'' in Euclidean toroidal compactifications:

\noindent
{\bf Proposition 7}: The set of compactification data
\eqn\cpdta{\eqalign{
E &=N^{-1} Diag\{R_1^2,\dots R_n^2\} (N^t)^{-1} + B \cr
R_i&\geq t^{-1} R_{i+1} \qquad 1\leq i\leq n-1\cr
R_{n-1} R_n&\geq t^{-1}\ , \qquad R_n \geq 1\cr
|n_{ij}|&\leq \half \ ,\qquad |B_{ij}| \leq \half\cr}
}
where $N$ is upper triangular and $t\geq {\sqrt{3}\over 2}$
contains a fundamental domain for the action of the
duality group on $\CB$.

\noindent
{\it Proof}: All the inequalities in \cpdta\ follow
by combining Proposition 6  and \efriwa, except for
$R_n\geq 1$. To establish this, note that
the fundamental domain described by $\Sigma_{t,\omega}$
is invariant under the permutation of simple roots
$\alpha_n\leftrightarrow \alpha_{n-1}$. This is just an
element of the Weyl group corresponding to the
exchange $x_n\leftrightarrow x_{2n}$ and right multplication
by it preserves the $KAN$ decomposition. Since it
takes $R_n \leftrightarrow R_n^{-1}$ we may fix it with
$R_n\geq 1$. $\spadesuit$

{\bf Remark}: Simultaneously bounding the radii in
toroidal compactification is {\it not} a trivial generalization
of the one-dimensional case. This appears
to have been first noted by E. Silverstein
\ref\silverstein{E. Silverstein, ``Large-Small
Equivalence in String Theory,'' hepth-9201009 }. She correctly
noted that one could either bound all the radii or
transform the volume to infinity by duality
transformations. Proposition 7 settles the matter in
favor of the first case.

Returning to the description of $\IL$ at infinity,
we characterize these regions in
terms of the eigenvalues $R_1^2,\dots R_n^2$ of the
metric $G$. Mathematically,
``regions of infinity'' in $\IL$ are described by
degenerations of sets of simple roots
\ref\bori{See \borel, p. 79, and section 12}.
That is,
a region of infinity is defined by specifying a
subset $\theta$ of simple roots and requiring
$\alpha_i <\epsilon$ for $\alpha_i\in \theta$.
Of course, the regions at infinity are connected in
complicated ways. The
maximal regions of infinity are defined by letting
one root approach zero holding all others fixed.
One may easily check that letting the spinor
root $\alpha_n=\lambda_{n-1} \lambda_n$ approach zero
holding all other
roots fixed is total decompactification, while the
analogous sets for other roots corresponds to
partial decompactification. Thus, in the
Euclidean case, the ``regions at infinity'' are in
complete accord with our common sense.

Although it is difficult to find a precise fundamental
domain for the duality group, the problem becomes
much more tractable near infinity.
Roughly speaking, the theorem of parabolic transformations
\ref\borii{See \borel\ Proposition 12.6}
characterizes those duality
transformations that are left unfixed at infinity.
More precisely, $\{\gamma : \Sigma_{t,\omega}\cdot \gamma\cap
\Sigma_{t,\omega}\not= \emptyset\}$ is finite and
lies in a certain ``maximal parabolic subgroup.''
For the case of the degeneration by the spinor root
(total decompactification) the maximal
parabolic is easily found to be:
\eqn\mxpara{
P_{\Delta-\{\alpha_n\}}=\{g:
g=\pmatrix{(A^{tr})^{-1}&0\cr
0&A\cr}\cdot \pmatrix{1& B\cr 0&1\cr}: A\in GL(n,\IZ), B_{ij}\in \IZ \}
}

The situation is far more complicated in the Minkowskian case
because we do not have the Iwasawa decomposition. No general
theory exists. We will merely
illustrate a few points with an example.

$\underline{Example: Decompactifying\ two \ dimensions}$

In both the Euclidean and Minkowskian cases
we begin with the space:
\eqn\elldmt{
\IL=\biggl[SL(2,\IR)/SL(2,\IZ)\times SL(2,\IR)/SL(2,\IZ)\biggr]/(\IZ_2\times
\IZ_2)}
In the Euclidean case compactifications are represented by
$SO(2)\times SO(2)$ orbits. The two ways to approach infinity
are
$$\eqalign{
\alpha_1\to 0: R_1/R_2 \to \infty &,  (R_1 R_2)<\infty\cr
\alpha_2\to 0: R_1/R_2 <\infty &,  (R_1 R_2)\to \infty\cr}
$$
One of the $\IZ_2$ subgroups in \elldmt\ corresponds to
the action of the Weyl group and identifies these boundaries.
Since the homomorphism
$\psi$ preserves Iwasawa decompositions the two ways to
go to infinity correspond to $Im \rho\to \infty$, $\tau$ fixed, and
$Im \tau\to \infty$, $\rho$ fixed. The Weyl group
action is the exhchange $\rho\leftrightarrow \tau$.

In the Minkowskian
case. Compactifications are represented by $SO(1,1)\times SO(1,1)$
orbits in \elldmt. These in turn may be thought of as pairs of
geodesics in $T_1\Sigma \times T_1\Sigma$, up to
$\IZ_2\times \IZ_2$ identifications. The orbits
extend to infinity in $\IL$ even for points which are -
in some sense - far from the uncompactified limit. For example,
the special compactification
$$\Gamma_* = (II^{1,1};0)\oplus (0;II^{1,1})$$
corresponds to the pair of geodesics specified by
$$(\rho_-,\rho_+,\tau_-,\tau_+)=(\infty,0,\infty,0)\qquad ,$$
which projects to the image of the
imaginary axis in each modular curve.
On the other hand, the compactification data
$$E=\pmatrix{-R_0^2&0\cr 0& R_1^2\cr}$$
corrsponds to the pair of geodesics with basepoints
$$(\rho_-,\rho_+,\tau_-,\tau_+)=(-R_0/R_1,+R_0/R_1,-R_0 R_1,+R_0 R_1)$$.
In particular, what is naively a ``very decompactified torus,''
with both $R_1R_0, R_0/R_1$ large can have an orbit which is
``close'' to the orbit of $\Gamma_*$, at least when projected
to $\Sigma$.

\subsec{Generalization to superstrings}

The above considerations can be generalized to
strings with $N=1,2$ local supersymmetries.

For $(1,1)$ supersymmetries
we let $h$ be the Loop algebra on
$\IR^{1,9}$.
We replace the symmetric algebra $S(h^-)$
by the corresponding ``Weil algebra''
$W(h^-)=S(h^-)\otimes \Lambda(h^-)$. Combining
left and right movers we have
for the type II superstring
\eqn\nsctr{
\CH_\Gamma^{NS}= W(h_L^-)_+\otimes
W(h_R^-)_+ \otimes \IC[\Gamma] }
where the subscript refers to a GSO projection.
The Ramond sector is given by
\eqn\rmdsctr{
\CH_\Gamma^{R}= W(h_L^-)_+\otimes
W(h_R^-)_+ \otimes \IC[\Gamma]\otimes \CS_+\otimes \CS_\pm}
where $\CS_\pm$ are the $16$-dimensional
Majorana-Weyl spinor representations of
$Spin(1,9)$. The $\pm$ distinguishes $IIA$ and
$IIB$ theories. The full matter statespace
associated to a lattice $\Gamma\in\CM$ is
$\CH_\Gamma=
\CH_\Gamma^{NS}\oplus\CH_\Gamma^{R}$.
One interesting subtlety arises in the
discussion of duality generalizing
\duality\
\ref\dnsb{M. Dine, P. Huet, and N. Seiberg, ``Large and
Small Radius in String Theory,'' Nucl.Phys.B322:301,1989.}.
The spinor representation $\CS_+\otimes \CS_\pm$
of $O(\eta)\times O(\eta)$ defines a
homogeneous vector bundle on $\IH$.
Right-multiplication by $A\in O(\check D;\IZ)$
which is not in the connected component
$O_0(\check D;\IZ)$ (e.g. left and right-moving parity)
may switch $\CS_+\leftrightarrow \CS_-$.
In such cases these components of the duality group
do {\it not} act as symmetries.

Similarly for $(2,2)$ supersymmetry we take
$h$ to be the loop algebra on $\IR^{2,2}$ and
form the Weil algebra as above
\ref\matmuk{S.D. Mathur and S. Mukhi, ``The $N=2$
Fermionic String,'' Nucl. Phys. {\bf B302}(1988)130}
\ref\ogvf{H. Ooguri and C. Vafa, ``Self-Duality
and $N=2$ String Magic,'' Mod.Phys.Lett. A5: 1389-1398, 1990; ``Geometry of
$N=2$ Strings,'' Nucl.Phys. B361: 469-518, 1991}. There is no need to
distinguish NS from R or any other twisted
sectors
\ref\schwsb{A. Schwimmer and N. Seiberg, Phys.
Lett. {\bf 184B}(1987)191}.

For heterotic strings with $(0,1)$
supersymmetries the only important change is
that we must define $O(D;\IR)$ and $O(\check D;\IR)$
by taking the direct sum of $D$ in \metric\ with the
metric tensor of an even self-dual Euclidean
lattice of dimension $16$. One can also
extend the discussion to heterotic
strings of type $(0,2)$ or $(1,2)$
using the ideas of
\ref\ogvfi{H. Ooguri and C. Vafa, ``$N=2$ Heterotic
Strings,'' Nucl.Phys.B367:83-104,199}.

\newsec{BRST Cohomology}

\subsec{Review}

The string theories associated to the
conformal field theories $\CH_\Gamma$, $\Gamma\in\IL$
are defined by the off-shell chain complex
\eqn\chainsp{
\CC_\Gamma^* =\CF_{b,c}\otimes \CF_{\bb,\cb}\otimes\CH_\Gamma
}
with BRST differential $Q+\bar{Q}$.
The on-shell space is the BRST cohomology $H^*(\CH_\Gamma)$.
Both $\CC^*$ and $H^*$ are bi-graded by left and right
ghost number: $H^{g,\bar{g}}$. We will work with absolute
cohomology for simplicity. Modifications for relative and
semirelative are straightforward.

The essential theorem we will use was established in
the proofs of the chiral no-ghost theorem
\nref\brower{R.C. Brower, ``Spectrum-generating algebra
and no-ghost theorem for the dual model,'' Phys. Rev.
{\bf D6}(1972)1655}
\nref\goddthrn{P. Goddard and C.B. Thorn, ``Compatibility
of the dual Pomeron with unitarity and the absence of ghosts in the dual
resonance model,'' Phys. Lett. {\bf 40B}(1972)}
\nref\freemnoliv{M.D. Freeman and D.I. Olive, ``BRS Cohomology
in String Theory and the No-Ghost Theorem,'' Phys. Lett.
{\bf 175B}(1986)151}
\nref\thorn{C.B. Thorn, ``Computing the Kac determinant using dual model
techniques and more about the no-ghost theorem,'' Nucl. Phys.
{\bf B248}(1984)551}
\nref\FGZ{I. Frenkel, H. Garland, and G. Zuckerman, ``Semi-infinite
cohomology and string theory,'' Proc. Nat. Acad. Sci. {\bf 83}(1986)8442}
\nref\GSW{M. Green, J. Schwarz, and E. Witten, {\it Superstring theory\/},
Cambridge Univ. Press (1987).}
\refs{\brower {--}\GSW}.
We will follow most closely the statement in \FGZ.

{\bf Theorem}\refs{\brower {--}\GSW}.
The absolute
chiral BRST cohomology, $H_A^*(\CF_p)$ is
nonzero only for $*=0,1,2,3$ and is given by
\eqn\zercoho{\eqalign{
H^{g=0}_A(\CF_p) &= \IR \cdot |1\rangle \qquad p=0\cr
&=0 \qquad\quad  p\not=0\cr
H^{g=1}_A(\CF_p) &= \IR^{26} \qquad p=0 \cr
&=\IR^d  \qquad d=p_{24}(n) \qquad p^2=2-2n, n\geq 0, p\not=0\cr
&=0 \qquad \quad {\rm otherwise}\cr}
}
together with $H_A^g\cong H_A^{3-g}$. Here $p_{24}(n)$
denotes the number of partitions into $24$ colors.

We will have occasion to use the equivalence to the
old physical state conditions. This establishes an
isomorphism of $H^{g=1}$ with the space of dimension
one Virasoro primaries modulo spurious states:
the ghost number one cohomology is spanned by elements
of the form $c V$ where $V$ is a representative of
\eqn\oldphys{
\CF_p[1]^{Vir^+}/(\CF_p[1]^{Vir^+}\cap Vir^-\cdot \CF_p)\qquad\qquad .
}
Here $Vir^\pm$ are the positively/negatively moded Virasoro
subalgebras and the notation $\CF_p[1]$ means restriction to
the space of dimension one operators.

Completely analogous results hold for the SuperVirasoro
algebra
\GSW
\ref\lzsv{B.H. Lian and G.J. Zuckerman, ``BRST Cohomology
of the Super-Virasoro Algebras,'' CMP {\bf 125}(1989)301}.
In particular, at ghost number $1$, where
the superfield $C=c+\theta \gamma$ has ghost number 1,
there is a one-one correspondence between matter primary
superfields of $h=1/2$ modulo spurious states,
and BRST invariant states. The discontinuity in \zercoho\
at $p=0$, which plays an important role below,
goes through with minor modifications. Analogous results
hold in the $N=2$ case
\ref\bien{Jadwiga Bienkowska, ``The generalized
no-ghost theorem for N=2 SUSY critical
strings,'' hep-th/9111047}.

\subsec{BRST Cohomology on $\IH$,$\IL$,$\CN$}

The absolute cohomology for \chainsp
is trivially calculated:
\eqn\clsd{
H^*(\CH_\Gamma)=\oplus_{(p_L;p_R)\in \Gamma} H^*_Q(\CF_{p_L})\otimes
H^*_{\bar{Q}}(\bar\CF_{p_R})
}

At a generic point
\foot{i.e., on a dense subset of total measure}
in $\IL$ the cohomology $H^*$
is {\it finite}dimensional, and localized at momentum zero.
This follows since by \zercoho\ $H^*(\CF_p)$ is zero unless
$p^2$ is an even integer $\leq 2$, but for generic lattices
$p_L^2,p_R^2$ are not integral unless $p_L=p_R=0$. Thus, at
generic points in $\IL$:
\eqn\gencoho{\eqalign{
H^1&=\oplus_{\mu} c\p X^\mu \cdot \IR
\oplus_{\mu} \cb\pb X^\mu \cdot \IR\cong \IR^{26}\oplus \IR^{26}\cr
H^{1,1}&=\oplus c\cb\p X^\mu\pb X^\nu\cdot\IR
\cong \IR^{26}\otimes \IR^{26}\cr}
}

\noindent
{\bf Remarks}:

1. The vast reduction of physical states noted above is
not a stringy phenomena. Consider the D'Alembertian
equation in a $1+1$-dimensional field theory in
which time and space dimensions are incommensurate.
The only solutions to the equations of motion are
constants.

2. It is natural to consider the
asymmetric orbifold
\ref\nsv{Narain, Sarmadi, and Vafa, ``Asymmetric
Orbifolds,'' Nucl.Phys.B288:551,1987}\
obtained by dividing by the $\IZ_2$ transformation
$Y\to -Y$ on both left and right-movers. In this case,
for generic lattices, there is {\it no} $H^{1,1}$
cohomology since the twisted sectors have
nonintegral $L_0,\bar{L}_0$. The entire
(absolute) cohomology is $(H^0+H^3)\otimes (\bar{H}^0+\bar{H}^3)$.

\subsec{Reduced String theories}

At less generic points in $\IL$ we can produce interesting
examples of string compactifications where the space of
physical states is infinite-dimensional, but still
far smaller than normally encountered in string theory.
In this section we sketch one such example.

Consider the point $E=Diag(-R_0^2, R_1^2,\dots R_{25}^2)$,
and the lattice
\eqn\prodcirc{\Gamma_{\vec R}=Span\{{1\over \sqrt{2}}
({n_{\mu}\over R_{\mu}}+m_\mu R_{\mu};
{n_{\mu}\over R_{\mu}}-m_\mu R_{\mu})  \}
}
which has generator matrix
\eqn\genmtrx{\CE=
exp\Bigl[\pmatrix{ 0 & \lambda\cr \lambda& 0\cr}\Bigr]
{1\over \sqrt{2}}\pmatrix{\eta& 1\cr \eta & -1\cr}
}
with $\lambda=Diag\{ \log R_{\mu}^{-1}\}$.

If the $R_{\mu}^2$ are all irrational
and incommensurate the cohomology is
described by \gencoho. If some spatial
dimensions, say $R_i$, $i=1,\dots d$
have a length commensurate with
the time direction then there will be an
infinite number of physical states. If $R_0^2$ is
rational we have an enhanced symmetry point
as described in the next section.

If $R_0^2$ is
irrational there are still an infinite number of
states but these will always occur at levels
$(0,2),(1,1),(2,0)$.
To see this, let $R_i=({p_i\over q_i}) R_0$ and suppose
$R_0^2$ is irrational and incommensurate with $R_j^2$ for $j>d$.
A state with momentum $(p_L;p_R)$ will be in the cohomology if
$p_L^2$, $p_R^2$ are even integers less than or equal to two.
Clearly this requires the components in the directions
$j>d$ to be zero. Moreover since $R_0^2$ is irrational
we must have
\eqn\cohconds{\eqalign{
 - n_0^2 + \sum_{i=1}^d n_i^2 ({p^i\over q^i})^2 &=0\cr
- m_0^2 + \sum_{i=1}^d m_i^2 ({q^i\over p^i})^2 &=0\cr}
}
Since $p_L^2$ and $p_R^2$ are both $\leq 2$ we have
$\pm n \eta m\leq 2\Rightarrow
n \eta m\in \{ -2,0,2\}$.

We can produce an infinite set of solutions to
\cohconds\ by generalizing the usual construction
of Pythagorean triples.
The states associated with these solutions clearly have levels
$(0,2),(1,1),(2,0)$. The $(1,1)$ are the familiar massless
particles of bosonic string. The $(0,2),(2,0)$ states have
energies rapidly going to infinity as we decompactify.
Curiously, there are no tachyons.

{\bf Remark}: The chaotic nature of the physical
states is reminiscent of the
nature of the special states
in the 2D string.

\newsec{Symmetries in Toroidal Compactification}

\subsec{Unbroken symmetries in string theory}

Gauge symmetries of CSFT are associated with
 G=1 BRST cohomology classes.
It is a well-known fact that the chiral ghost number
one cohomology defines a Lie algebra. It has
been discussed, for example, in
\nref\graeme{G. Segal, Lectures at the Isaac Newton Institute,
August 1992, and lectures at Yale University, March 1993.}
\nref\lziii{B. Lian and G. Zuckerman,
``New perspectives on the brst algebraic structure of string theory,''
(hep-th/9211072) Toronto-9211072.}
\nref\getzler{E. Getzler, ``Batalin-Vilkovisky Algebras and
Two-Dimensional Toplogical Field Theories'' (hep-th/9212043)}
\FGZ\refs{\graeme{--}\getzler}\
and elsewhere.
Similarly, for the superstrings we again look
for $G=1$ cohomology. The superline integrals
$\oint dz d\theta J$ where $J$ is a weight $1/2$
superfield form a Lie superalgebra.

\subsec{Enhanced symmetries}

{\bf Definition}. A point $\Gamma\in \IL$ is called
an {\it enhanced symmetry point} (ESP) if $\Gamma$ contains a
positive rank sublattice of the form
$$(\gamma_L;0)\oplus (0;\gamma_R)\subset\Gamma$$
where $\gamma_L$ or $\gamma_R$ contains vectors of
norm $\leq 2$.

At enhanced symmetry points there will be extra
ghost number one cohomology classes. These will
always be of the form $cJ(z)$ or $\bar{c}\bar{J}(\zb)$
and form a Lie algebra
$\CL_\Gamma=H^{0,1}\oplus H^{1,0}$.
The set of enhanced symmetry points contains the
$O(\eta)\times O(\eta)$ orbit of $O(\check D;\IQ)/ O(\check D;\IZ)$
and is a dense set of measure zero.

{\bf Example 1}. At a generic point
$\CL_\Gamma=\IR^{26}\oplus \IR^{26}$.

{\bf Example 2}.  $\gamma_L$ is  a root
lattice of a simply laced group. $H^{1,0}$
is a finite dimensional Lie
algebra.

{\bf Example 3}. Generalized Kac-Moody Algebras
\ref\borcherds{R. Borcherds, ``Generalized Kac-Moody
Algebras,'' J. Algebra {\bf 115}(1988)501; ``Monstrous
moonshine and monstrous lie superalgebras,''
Invent. Math. {\bf 109}(1992)405}.
When the lattice $\gamma_L$ is hyperbolic we obtain
infinite dimensional symmetries of the ground state.
These are already quite nontrivial if the ESP
lattice $\gamma_{L,R}$ is $II^{1,1}$ or $II^{1,1}\oplus \sqrt{2}\IZ$.
In the latter case a related hyperbolic algebra
has been extensively studied
\ref\feinfren{A.J. Feingold and I.B. Frenkel, ``A hyperbolic
Kac-Moody algebra and the theory of Siegel modular
forms of genus 2,'' Math. Ann. {\bf 263}(1983)87}.

\subsec{A distinguished compactification}

{\bf Proposition 8}.
If a timelike dimension is compactified then
there is a unique point
$\Gamma_*\in \CN$ in the Narain moduli
space at which the closed string
completely factorizes between left and right movers:
\eqn\fctrthry{
\CH_{\Gamma_*}=\CC \otimes \bar{\CC}
}
This is true in any string theory with $N=0,1,2$
supersymmetries.

\noindent
{\it Proof}. The theory factorizes iff
the lattice $\Gamma$ may be written as
a direct sum of left- and right-moving lattices
\eqn\maxlattic{
\Gamma=(\Gamma_L;0)\oplus (0;\Gamma_R)
}
Since $\Gamma$ is even integral  so are $\Gamma_{L,R}$.
Since $\Gamma$ is unimodular it follows that $\Gamma_{L,R}$
are too. We now apply the uniqueness of even unimodular lattices
with noneuclidean signature. For example, for
the totally compactified $N=0,1,2$ strings we have:
\eqn\maxpoint{\eqalign{
\Gamma_* &\equiv (II^{1,25};0)\oplus (0;II^{1,25})\cr
\Gamma_* &\equiv (II^{1,9};0)\oplus (0;II^{1,9})\cr
\Gamma_* &\equiv (II^{2,2};0)\oplus (0;II^{2,2})\cr}
}
respectively,
where $II^{p,q}$ is the unique even integral unimodular lattice
in $\IR^{p,q}$. We thus obtain \fctrthry\ with
\eqn\univopen{
\CC=S(h_{1,25}^-)\otimes \IC[II^{1,25}]}
for the bosonic string, and similarly for
the $N=1,2$ strings. Similar remarks also apply to
heterotic strings. $\spadesuit$

{\bf Remarks}:

1. The distinguished point $\Gamma_*$ may be
regarded as a point of maximal symmetry in the moduli space of
toroidal compactifications in the following sense:
At an enhanced symmetry point the Lie algebra of
symmetries $H^1=H^{1,0}\oplus H^{0,1}$ organizes the rest of
the statespace into representations. For example,
the space of ghost number $1,1$ states may be decomposed as
\eqn\ghnmtw{
H^{1,1}=H^{0,1}\otimes H^{1,0} \oplus_i H_i\otimes \bar{H}_i
}
where $H_i,\bar{H}_i$ are representations of $H^{0,1},H^{1,0}$.
If the rank of $\gamma_L,\gamma_R$ is maximal then the
sum will be finite and the theory is rational,
otherwise it is infinite and the theory is quasirational.

In CSFT zero-modes of
``gauge bosons'' are of the form
$J_1 \bar J_2\in H^{0,1}\otimes H^{1,0}$,
that is, we should regard gauge bosons as the
part of the ghost number 2 cohomology whose
existence is dictated by the symmetry $H^{0,1}\oplus H^{1,0}$.
At the point $\Gamma_*$ only the adjoint representation
appears: ``All states are (zero-modes of) gauge bosons.''
In general we will refer to states in $H^{0,1}\otimes H^{1,0}$
as ``symmetry states.''

2. Given a point of maximal symmetry it is natural to
ask if $\CL_*=\CL_{\Gamma_*}$ is a universal symmetry
of string theory in the sense that all
other unbroken symmetry algebras which arise
in toroidal compactification are subalgebras of
$\CL_*$.
Unfortunately, maximal symmetry does not imply
that $\CL_*$ is universal.
 To see that $H^1$ is not universal note
that the only way in which we could have a Lie algebra
embedding would be to have $(\gamma_L;0)\oplus (0;\gamma_R)$
be a sublattice of $\Gamma_*$. Consider the case of
maximal rank. Then $(\gamma_L;0)\oplus (0;\gamma_R)$
is only a sublattice if $\det G$ for its metric
tensor $G$ is a perfect square. It is easy, using the
examples of section 3.3, to construct ESP's which do
not have this property.

3. For the bosonic string the Lie algebra $\CL_*$
is related to the Monster group.
$\CL_{*}=
\CA \times \CA$ where $\CA$ is
the ``Fake Monster Lie algebra'' studied by
Borcherds \borcherds. For $(1,1)$ supersymmetry
$\CA$ is related to the hyperbolic Kac-Moody
algebra $E_{10}$.

4. A natural
geometric construction of the  groups corresponding
to $\CL_*$ would be
very interesting.


\subsec{Lorentz orbits and duality orbits}

The ergodic nature of the duality action on
$\IH$ and of the Lorentz action on $\IL$ discussed
in section 2.5 raises the issue of the nature of
the orbits of ESP's, especially of the distinguished
orbit $\Gamma_*$.  Mathematically this appears to be a difficult
and deep problem, but physical common sense leads us to expect that
an enhanced symmetry point cannot have a dense orbit.
The reason is that ESP's correspond to {\it quantum} symmetries
arising from special configurations of the spacetime lattice.
If the Lorentz orbit of an ESP were dense then we could decompactify the
spacetime torus and, no matter how large we make all the spacetime
radii, there would be a nearby ESP. This strongly
contradicts the physical idea that as one decompactifies
one must recover the physics of the uncompactified theory.
Therefore we state a:

{\it Conjecture}: Let $\Gamma$ be an ESP. Then the
$O(\eta)\times O(\eta)$ orbit of $\Gamma\in \IL$ is closed in
$\IL$.

{\bf Example}: Consider compactification of $1+1$ dimensions as
in section 2.6 above.
If $E$ is rational then $\rho_\pm, \tau_\pm$
solve a quadratic equation with rational coefficients. Therefore
the continued fraction expansions are periodic
\ref\kengii{See \keng, Theorem 6.1}.
{}From the considerations of section 2.6 we see that the orbit is
closed and not dense.

We can also settle the question for the particular case of
the orbit $\Gamma_*$.

\noindent
{\bf Proposition 9}. The Lorentz orbit $\Gamma_*\subset \IL$ is
a closed $O(\eta)\times O(\eta)$ orbit.

\noindent
{\it Proof}. Consider the space $\hat\CM$ of all generator
matrices for lattices $\Gamma\in\IL$. Thus, we consider
matrices $\CE$ such that $\CE^{tr}D\CE$ is integral, unimodular,
with even diagonal. We have $\IL=\hat\CM/GL(52;\IZ)$ and
$\hat\CM$ admits a left $O(D;\IR)$ action. For a
point $\CE_*$ in the orbit $\Gamma_*$ we can find an
integral change of basis $A\in GL(52;\IZ)$ such that
\eqn\chgbsis{
\CE_*\cdot A = \pmatrix{ * & 0 \cr 0& *\cr}
}
Moreover, the left $O(\eta)\times O(\eta)$ action
preserves this form. Thus, it suffices to show that there
are generator matrices $\CE\in\CM$ which cannot be brought
arbitrarily close to this form by right action of $GL(52;\IZ)$.
Write
$$\CE=\pmatrix{\CE_{11}& \CE_{12}\cr \CE_{21}& \CE_{22}\cr}
$$
and assume $\CE_{11}$ is invertible. Then the equations
$(\CE\cdot A)_{12}=(\CE\cdot A)_{21}=0$ are equivalent to
\eqn\blckzr{\eqalign{
A_{12}+\CE_{11}^{-1}\CE_{12} A_{22}&=0\cr
A_{11}+\CE_{21}^{-1}\CE_{22} A_{21}&=0\cr}
}
Choosing $\CE_{11}^{-1}\CE_{12}=E=Diag\{ R_{\mu}^2\}$ with
$R_{\mu}^2=p_{\mu}/q_{\mu}$ rational we may take $\CE$ to
be given by \genmtrx.
It is then easy to see that the most
general integral matrix $A$ solving \blckzr\ is
\eqn\slvblck{
A=\pmatrix{Diag\{ p_{\mu}\} & 0\cr 0 & Diag\{ q_{\mu}\}\cr}
\pmatrix{1&1\cr 1&-1\cr} \pmatrix{\ell_1&0\cr 0& \ell_2\cr}
}
where $\ell_{1,2}$ are invertible, integral, $26\times 26$
matrices. Then
\eqn\slvblcki{
|\det A|=\biggl| 2^{26}\det \ell_1 \det\ell_2 \prod_{\mu}
(p_{\mu} q_{\mu}) \biggr|
}
since any matrix $A\in GL(52;\IZ)$ always has $|det A|=1$,
while \slvblcki\ is bounded away from $1$ we see that no
$GL(52;\IZ)$ matrix can make both
$(\CE\cdot A)_{12}$ and $(\CE\cdot A)_{21}=0$ arbitrarily
small. $\spadesuit$

\subsec{Symmetry Points and Orbifold Points}

The action of the duality group on $\IH$
sometimes has nontrivial fixed points.
In the Euclidean case these correspond to orbifold points of
$$\CN=O(n)\times O(n)\backslash \CM/O(\check D;\IZ)$$
{}From the point of view of right orbits of the duality
group on $\IH$ this comes about when elements of the
duality group leave $E$ fixed. From the point of view of left orbits of
$\IL$ these orbifold points correspond to lattices $\Gamma\in\IL$ which have
automorphism groups $Aut(\Gamma)$ containing elements
other than the trivial automorphism $x\to -x$.

This discussion generalizes straightforwardly to the
Minkowskian case. However,
it often happens that the automorphism groups of hyperbolic
lattices are in fact {\it infinite}
discrete groups, so the singularities at the ``orbifold
points of $\CN$'' can be rather wild. A spectacular example of
this is provided by the orbits corresponding to
$\Gamma_*$. For example, the fixed subgroup of the
duality group is the infinite discrete
group $Aut(II^{1,25})\times Aut(II^{1,25})$.
For an interesting description of this group see
\conway\ ch. 27

Orbifold points and enhanced symmetry points appear
to be closely connected. Indeed, if the automorphisms
are of $\Gamma\in\IL$ are of the form $(R,1)$ or $(1,R)$
then the orbifold point is necessarily an enhanced symmetry point,
since $(R p_L,p_R)-(p_L;p_R)=(p_L';0)$ is in $\Gamma$ with
$p_L'\not=0$. Conversely,
enhanced symmetry points related to affine Lie algebras
or points with vectors of square-length two automatically
correspond to orbifold points. The reason is that the
Weyl reflection in a vector of length two defines a nontrivial
automorphism of the lattice.

For these reasons orbifold points are often identified
with ESP's. We have not managed to prove this.
Because of the possibility that the automorphism
group is nontrivially embedded in $O(\eta)\times O(\eta)$
we must ask:

{\it Question 1}: Are all orbifold points enhanced
symmetry points?

Conversely, there are enhanced symmetry points
containing no vectors of length two. In all cases we have
examined these ESP's are also orbifold points so we also
ask:

{\it Question 2}: Are all enhanced symmetry points orbifold
points?

\newsec{Spontaneous Symmetry Breaking}

\subsec{Disappearing states}

Toroidal compactification of all dimensions presents an
interesting example of backgrounds in which there are large unbroken
symmetries together with infinite-dimensional BRST cohomology.
When we move away from these points in a generic direction
the symmetries disappear. Indeed,
there is a standard description of
spontaneous symmetry breaking in toroidal compactification
\GSW.
The values of the couplings to the exactly marginal
operators $\p Y^a \pb Y^{\bar a}$ in the Lagrangian
correspond (up to contact term subtleties) to the
vacuum expectation values of exactly massless
spacetime
scalar fields $\Psi^{a,\bar{a}}$. ESP's correspond
to points where the vacuum expectation values leave
unbroken gauge symmetries larger than
$U(1)^d\times U(1)^d$.

A generic perturbation away from an ESP such
as $\Gamma_*$ has drastic consequences - an infinite
number of states disappears! While initially
surprising, this is quite in line with the standard
interpretation of spontaneous symmetry breaking
in toroidal backgrounds.
Consider  the string field theory action
for in an ESP background $\Gamma$.
Calculation of the action of \bz\
will result in an expression like
\eqn\sftact{
S(\Psi)=\sum_{(p_L;p_R)\in \Gamma} \Psi_I(p_L,p_R) (Q^{IJ}-(m^2)^{IJ})
\Psi_J(p_L,p_R)
+
\sum_{p_1,p_2,p_3} K^{IJ}_{a,\bar{a}} \Psi^{a,\bar{a}}\Psi_I \Psi_J +\cdots
}
where $K$ is some typically nonvanishing coupling
computed from the Witten three-string vertex. The
ellipsis indicates the higher terms of the nonpolynomial
theory. $\Psi^{a,\bar{a}}(p_L,p_R)$ are spacetime
fields corresponding to the exactly marginal directions
$\p Y^a \pb Y^{\bar a}$.

The masses $(m^2)^{IJ}$ in \sftact\
are integrally quantized while the
momenta are constrained to lie on a lattice.
Therefore, if we shift the vacuum expectation value of
the fields $\Psi^{a,\bar{a}}(0,0)$ the shift in the
three-point interactions modifies the quadratic terms
and in general leaves no zero-modes
for the perturbed quadratic form
acting on the lattice.

Thus, although the absence of states at a generic
point is perfectly sensible from the spacetime
field theory point of view as explained in section
3.2 the main novelty in string theory is that the global
dimensions of space and time can be modified by
giving vacuum expectation values to the
spacetime fields. That is, the disappearance of states
is the result of spontaneous symmetry breaking.

\subsec{Goldstone vs. Higgs}

The Higgs
mechanism seems to be at odds with the discontinuity
in the number of on-shell degrees of freedom
noted above. Let us clarify this point.

Suppose there are $d$ compact Euclidean
directions and $26-d$ noncompact directions. Furthermore
consider a family of compactifications $g(t)\cdot \Gamma$
where $g(t)$ is a family of $O(D)$ Lorentz transformations
and $\Gamma$ is an ESP. Consider the
evolution of the cohomology groups which at $t=0$
correspond to the gauge bosons of the enhanced symmetry.
Let
\eqn\dimgrps{d(t)=
\dim H^{1,1}(\CF_{p_L(t)}\otimes \bar{\CF}_{p_R(t)}
\otimes \CF_{q(t)}\otimes \bar{\CF}_{q(t)} )
}
where $q(t)$ denotes the momenta in the noncompact
directions, and $p_L(t=0)^2=2$, $p_R(t=0)=0$. Note that
$q(0)^2=0$, appropriate to a massless particle.

Of course, $d(t)$ is a constant function for $t\not=0$.
This follows from  section 3.1: time will
dress up compact momenta appropriately.

If $q(0)\not=0$ then $d(t)$ is continuous at $t=0$.
This statement of continuity of the number of
on-shell degrees of freedom is essentially a statement
of the Higgs mechanism.
The evolution to nonzero values of $t$
\foot{The ``evolution'' is defined precisely in
the section seven in terms of Lorentz transport.}
of the would-be scalar Goldstone mode
$$\delta_J (\p Y^a \Delta E_{a \bar{b}} \pb Y^{\bar b})
e^{i q(0)\cdot Z}
=e^{i p_L(0) Y} p_L(0)^a \Delta E_{a \bar{b}} \pb Y^{\bar b}
e^{i q(0)\cdot Z}, $$
where $Z$ denotes noncompact dimensions, is $Q$-equivalent
to the longitudinal mode of a massive vector boson:
$$   \epsilon_L\cdot\pb Z
e^{i p_L(t)Y} e^{i p_R(t)\bar{Y}} e^{i q(t)\cdot Z}$$
Put another way, the massless cohomology
of the rightmoving sector is, of course, 24 dimensional.
At $t=0$ we account for the degrees of
freedom this way:
$$24=d + (24-d)\qquad ,$$
for $d$ scalar modes and
$24-d$ massless vector modes in $26-d$ dimensions.
At $t>0$, we write:
$$24=(d-1)+(25-d)$$
since there are
only $d-1$ scalar modes and $25-d$ massive vector
modes in $26-d$ dimensions.

If $q(0)=0$ then $d(t)$ is discontinuous at $t=0$.
This statement of discontinuity of the number of
on-shell degrees of freedom is essentially a statement
of Goldstone's theorem. The discontinuity comes from the
massless right-moving cohomology. At $t=0$ all
rightmoving momenta vanish, so $H^1(\bar{\CF}_0)=26$.
Thus, at $t=0$ we have
$$26=d+ (26-d)$$
for the
zeromodes of $d$ massless scalars and the zero modes
of $26-d$ gauge bosons in $26-d$ dimensions.
At $t>0$ $d\to d-1$: we lose a scalar, and
$26-d\to 25-d$, the vector boson gets massive
and has nonzero momenta. This accounts for a loss
of {\it two} degrees of freedom.

When we toroidally compactify time
the infinite reduction in the number of
on-shell degrees of freedom in the neighborhood
of an ESP is merely an extreme example of the
the loss of the on-shell degrees of freedom
common to all theories of
spontaneously broken gauge theory.

\noindent
{\bf Remarks}:

1. The above discussion is closely related
to an elegant remark of E. Verlinde
\ref\vermster{E. Verlinde, ``The master equation of
2D string theory,'' hep-th/9202021; Nucl. Phys. B381 (1992) 141.}.
If a change of $\sigma$-model action
$\Delta S$ breaks a symmetry for a current
$J$ then $\delta_J \Delta S$, which would be
the zero-mode of the Goldstone boson, is always
$Q$-trivial, at least to first order in $\Delta S$.
It is not true that $\delta_J \Delta S$ remains
$Q$-trivial for finite deformations $\Delta S$.
In the case of toroidal compactifications
one can interpret Proposition 16 below as an
extension to all-orders in $\Delta S$
of Verlinde's remark.

2. There is one other way in which the string
example of SSB is strikingly different from
its field-theoretic counterpart. In the latter
case there is usually at least one point in
the space of Higgs vevs where the entire
gauge symmetry remains unbroken. Remark 2
of section 4.3 shows that  this
does not hold in the string case.
Nevertheless, we can try to construct such a
``universal symmetry.'' This is the subject of
the next section.

\newsec{A universal symmetry}

\noindent
{\bf Definition}: A universal symmetry $\CL$ is a Lie algebra
such that, for all ESP's $\Gamma\in \IL$ there is a
Lie algebra embedding $H^1(\CH_\Gamma)\hookrightarrow \CL$.

In this section we attempt to give a {\it natural}
construction of such a universal Lie algebra for the ESP's of
the toroidally compactified bosonic string.
\foot{It is easy to construct unnatural universal symmetries.
For example, just take the direct sum over all ESP's!}

We begin by unifying left- and right-moving oscillators
into a single Lorentz multiplet.
Note that instead of \defosc\ we could define
\eqn\fifosc{\eqalign{
\rho_m^A&=\beta_m^A \qquad \quad \quad A=1,\dots n+1\cr
&=\bar\beta_{-m}^{A-n-1} \qquad A=n+2,\dots 2n+2\cr}
}
and write the single equation
\eqn\fifcom{
[\rho_n^A, \rho_m^B]=D^{AB} n \delta_{n+m,0}\qquad \qquad .
}
This suggests that we should consider
the toroidal compactification of the {\it open string}
in a $52$ dimensional spacetime with the metric $D$.
The torus is obtained from the unique
even self-dual lattice $II^{26,26}$.
We use the quadratic form $D$ to
form a Heisenberg algebra from the loop algebra on
$\IR^{26,26}$ with polarization $h=h^-\oplus h^0\oplus h^+$, and form
\eqn\bighili{\CH=S(h^-)\otimes \IC[II^{26,26}]\qquad .}
Since $O(D)$ is transitive on $\IL$
one might think the appropriate Lie algebra is:
\eqn\univali{
\CL {\buildrel ?\over =} \CH[1]^{Vir^+}/(\CH[1]^{Vir^+}\cap Vir^-\CH)
}
Unfortunately, this Lie algebra does not quite work.

The problem is that the Lie algebra \univali\
only contains
$H^{1,0}\times (H^{0,1})^{opp}$
where $(H^{0,1})^{opp}$ is the {\it opposite}
Lie algebra to $H^{0,1}$
obtained by the following construction.
If $\CL_\Gamma$ is the Lie algebra associated to an
even integral lattice we may form the opposite
Lie algebra $\CL_\Gamma^{opp}$ by changing
$T\to -T$, $[\beta,\beta]\to -[\beta,\beta]$.
The definition of the vacuum remains unchanged, and
$\beta_{-n}$ still raises $L_0$ by $+n$ units for
$n>0$, but the sign of the zero-mode Hamiltonian
has changed: $H=-\half p^2 +N$ where $N$ is the
level. Thus, the roots of the Lie algebra in general
lie {\it outside} the lightcone, rather than
inside the lightcone. To obtain a universal
Lie algebra we need an algebra which is invariant under
the opposite mapping.

To remedy this problem we allow both purely real and
purely imaginary momenta
\foot{There is an analogy to including both
microscopic and macroscopic states, in Liouville
terminology}. Closure of the OPE then forces us to
consider linear combinations, but only with
integer coefficients. Therefore we consider an
extension of scalars from the $\IZ$ module $II^{26,26}$
to a module over the Gaussian integers:
\eqn\extsc{\tilde{\Gamma}^{26,26} = \IZ[i]\otimes_{\IZ} II^{26,26}
}
we now form
\eqn\bighilii{\tilde \CH=S(h^-)\otimes \IC[\tilde{\Gamma}^{26,26}]}
and denote by $ \CL_U $ the corresponding
Lie algebra of dimension 1 primaries
\eqn\unvslsymm{\CL_U\equiv
\tilde\CH[1]^{Vir^+}/(\tilde\CH[1]^{Vir^+}\cap Vir^-\cdot \tilde\CH)
}

{\bf Proposition 10}: All enhanced symmetries $H^{1,0}(\CH_\Gamma)\oplus
H^{0,1}(\CH_\Gamma)$ for $\Gamma\in\IL$ are Lie subalgebras
of $ \CL_U$.

{\it Proof}:
Using the projection $(\pi_L,\pi_R):\IR^{52}\to
\IR^{26}\oplus \IR^{26}$ any closed string theory statespace
is naturally a subspace of \bighilii\ where leftmoving oscillators are
interpreted
as the first 26 dimensions
\eqn\univii{
\prod (\beta_{n}^a)^k \prod (\bar{\beta}_{\bar n}^{\bar{a} })^{\bar{k}}
|p_L\rangle;p_R\rangle
\mapsto
\prod (\rho_{n}^a)^k \prod (i \rho_{\bar n}^{\bar{a}+26})^{\bar{k}}
|p_L; i p_R\rangle
}
Since $\CL_U$ is $O(D;\IR)$ invariant we can
``position'' the lattice so that the purely holomorphic or
antiholomorphic vectors of $\CH_\Gamma$ map into the first
or second $26$-dimensional components.
This mapping takes the ghost number one cohomology of a
closed string state into the Lie algebra
$\CL_U$ of $\tilde \CH$. Moreover, it preserves
the operator product structure for purely holomorphic or
anti-holomorphic states and hence defines a Lie algebra
embedding. $\spadesuit$

{\bf Remarks}:

1. In a sense which is difficult to make precise all of
the toroidally compactified closed strings should perhaps
be viewed as broken phases of the single theory $\tilde\CH$.
In particular, any given background breaks most of the
$ \CL_U$ symmetries.

2. The distinguished closed string and the ``open string''
defined by \bighili\ are closedly related.
Consider the conformal field
theory $\CC=S(h_{1,25}^-)\otimes \IC[II^{1,25}]$, which we may
refer to as the ``universal open string.'' We may combine
two copies of this theory in two distinct ways. First,
the holomorphic CFT $\CC$ admits an automorphism $I$ comming
from $z\to 1/\bar{z}$, i.e., time reversal symmetry on the worldsheet.
Alternatively, we may use the ``opposite map'' described above.
Then
\eqn\twthries{\eqalign{
{\rm ``52-dimensional\ open\ string"}\quad \CH &= \CC \otimes \CC^{opp}\cr
{\rm distinguished\ closed\ string}&=\CC \otimes I[\CC]\cr}
}
This observation leads to some obvious speculations. See the
conclusions.

3. The above construction of a universal symmetry is closely
related to some considerations of Giveon and Porratti
on duality-invariant effective actions
\ref\giveon{A. Giveon and M. Porratti,
``Duality invariant string algebra and $D=4$ effective actions''
Nucl.Phys.B355:422-454,1991;
``A completely duality invariant effective action of $N=4$ heterotic
strings,'' Phys.Lett.B246:54-60,1990}.
These authors proposed that similar large
(generalized) Kac-Moody symmetries
play a role as fundamental symmetries in toroidal compactification.
Indeed, if our construction is generalized to the heterotic
string in the obvious way our algebra would contain theirs as a
subalgebra. More recently, Giveon and Shapere have suggested a
candidate universal {\it group} for the case of the $N=2$ string
\ref\givshap{A. Giveon and A. Shapere, ``Gauge symmetries
of the $N=2$ string,'' hep-th/9203008;
Nucl. Phys. {\bf B386}(1992)43.}.

4. A theorem of Lian and Zuckerman
\ref\lziv{Theorem 3.5 in,
B.H. Lian and G.J. Zuckerman, ``BRST Cohomology and
Highest Weight Vectors I,'' Commun. Math. Phys. 135 (1991) 547-580}
states that any space
formulated in terms of the ``old physical state conditions''
such as \unvslsymm\ can be formulated in terms of BRST
cohomology. Thus it might be fruitful to examine the
(rather odd) corresponding $c=26$ open string theory.

5. As with the distinguished compactification symmetries, it would be
very interesting to find a natural geometrical interpretation
of the group corresponding to \unvslsymm.

6. Duality symmetries have been interpreted as being part of
some mysterious gauge group. This was first suggested in
\dnsb\ and further elaborated upon in \givgen. Indeed, at
ESP's producing affine Lie algebras the fixed subgroup
of the duality group may be interpreted as the Weyl
group of the gauge group. If remark 5  can be
understood it would be natural to expect that $O(\check D;\IZ)$
is some kind of Weyl subgroup.

\newsec{Relating points on $\IL$ and $\IH$}

We would like to study symmetry-breaking as a function of
the scalar field vev's, which define different backgrounds
$E\in\CB$ and hence different points on $\CN$. Therefore
we need a notion of parallel transport of theories.

\subsec{Connection and transport for a Lorentz-rotated Family}

In this section we will define a natural connection on the
Hilbert bundle $\CH\to \IL$  with fiber $\CH_\Gamma$. Indeed
this connection exists on any family of conformal field
theories with statespaces $\CH_{g\cdot \Gamma}$ of the form
\statspi\ or \statespace, where
$\Gamma$ is a lattice and $g$ belongs to a family of orthogonal
transformations.

Since a connection is not tensorial, to define it we
must choose a
local framing. Therefore consider a patch $U\subset O(D;\IR)$
and the family of theories $\CH_{g\cdot \Gamma}$ for
$g\in U$.
We choose our local framing of $\CH$ to be the set of states
\eqn\frame{
\prod \rho_n^A |g\cdot(p_L;p_R)\rangle
}
where $(p_L;p_R)$ range over the lattice $\Gamma$,
$\rho_n^A$ are defined in \fifosc\ and the products
of creation operators have $n<0$ for $A\leq 26$ and $n>0$ for
$A>26$.

In terms of $\rho_n^A$ we can define a representation of
$Lie(O(D;\IR))$:
\eqn\repod{
m_{AB}\in Lie(O(D;\IR)) \to \CO(m)\equiv \sum_{n\not= 0}{1\over 2n}
\rho_{-n}^A m_{AB} \rho_n^B
}
For $A,B$=$1,\dots 26$ or $27,\dots 52$ these are the
standard generators of left and right Lorentz transformations.
Here leftmovers and rightmovers get mixed by
a generic transformation.

The operators $\CO(m)$ are rather delicate.
They are not defined on the whole statespace,
so strictly speaking the Bogoliubov transformation does not make sense.
For example, $\CO(m)^2$ is an ill-defined operator, and we
cannot exponentiate $\CO(m)$. Nevertheless $Ad \CO(m)$ is well
defined acting on $S(h_L^-)\otimes S(h_R^-)$.

{\bf Definition}:
Let $m\in Lie(O(D;\IR))$ be a tangent vector to $\CM$
then the connection $\nabla_{m}:\CH_{g\cdot \Gamma}
\rightarrow\CH_{g\cdot \Gamma}$
is defined with respect to the frame \frame\ by
\eqn\defconn{
\nabla_{m}\prod \beta_n^A |g\cdot(p_L;p_R)\rangle
\equiv [\CO(m),\prod \beta_n^A ]|g\cdot(p_L;p_R)\rangle
}

\noindent
{\bf Proposition 11}. The connection $\nabla$ is flat.

\noindent
{\it Proof}: This follows because we have a representation:
$[\CO(m_1),\CO(m_2)]-\CO([m_1,m_2])=0$ implies
$[\nabla_{m_1},\nabla_{m_2}]-\nabla_{[m_1,m_2]}=0$. $\spadesuit$

The formula \defconn\ is easily exponentiated to give the
formula for parallel transport on $\IL$ by any
$g\in O(D;\IR)$:

\noindent
{\bf Proposition 12}: On a Lorentz-rotated family the parallel
transport under the connection $\nabla$
\eqn\trnsprti{
T^{g}:\CH_{ \Gamma}\to \CH_{g\cdot \Gamma}
}
is obtained from
\eqn\fnttrns{\eqalign{
\rho_n^A &\to \rho_n^{A'} (D g D)_{A'}^{\ A}\cr
|p_L;p_R\rangle &\to |g\cdot (p_L;p_R)\rangle\cr}
}

\noindent
{\it Proof}: One need only check that $Ad[\CO(m)]$ has
been properly exponentiated. Note that one must normal
order the final result after rotating the oscillators. $\spadesuit$

{\bf Example}: Consider the one-dimensional case with
\eqn\boost{
g(\lambda)=\pmatrix{\cosh \lambda& \sinh \lambda\cr
\sinh \lambda & \cosh \lambda\cr} }
then
\eqn\expltr{
T^g(\beta_{-n}\bar{\beta}_{-n}|0\rangle)=
\cosh^2\lambda\ \beta_{-n}\bar{\beta}_{-n}|0\rangle
+n \cosh \lambda \sinh \lambda\ |0\rangle
}
In particular, note that the space of exactly
marginal operators is {\it not} preserved by
the transport.

{\bf Remarks}:

1. The above construction gives a globally defined flat
connection on the vector bundles
$\CH\to \CM$ and $\CH\to \IL$. By choosing
coordinates one can also use it to construct a flat connection
on $\CH\to\IH$. This will be globally defined for
Euclidean compactifications, but only defined on an
open subset of $\IH$ in the Minkowskian case.

2. Similar formulae to those above have already
appeared in two papers by Kugo and Zwiebach
\ref\kugozwieb{Taichiro Kugo and Barton Zwiebach,
``Target Space Duality as a Symmetry of String Field Theory,''
hep-th/9201040
}
and by Ranganathan
\ref\rangan{K.Ranganathan, ``Nearby CFT's in the operator formalism:
The role of a connection,'' hep-th/9210090}.
Indeed our treatment was in part motivated by
these papers. Nevertheless, we believe our connection
differs from that in \rangan, as well as from
that in
\ref\rsz{K. Ranganathan, H. Sonoda, and B. Zwiebach,
``Connections on the State-Space over Conformal Field
Theories,'' hep-th/9304053}.
The main difference lies in the treatment of
the zero-modes. Strictly speaking,
\trnsprti\fnttrns\ is {\it not} a
Bogoliubov transformation, because $\CO(m)$
cannot be exponentiated. The parallel transport of
\rangan\ analogous to \expltr\ would produce a
state with arbitrarily high particle number
\foot{Compare, e.g., eq. (3.37) or (4.27) of
\kugozwieb\ or eq. (12,13) of \rangan.}
in contrast to our transport \expltr.

3. Cecotti and Vafa
\ref\cecvafa{S. Cecotti and C. Vafa, ``Topological anti-topological
fusion,'' Nucl. Phys. B367:359-461, 1991}
have defined a notion of parallel transport on the
vector bundle of chiral primary fields over moduli
spaces of $N=2$ theories. Since the
bosonic string can be regarded as a twisted
$N=2$ theory
\ref\semikhat{B. Gato-Rivera and A. Semikhatov, hep-th/9207004,
Phys.Lett.B293:72-80,1992}
\ref\lerche{M. Bershadsky, W. Lerche, D. Nemeschansky, and
N.P. Warner, hep-th/9211040},
one might therefore try to apply the results of
\cecvafa. The chaotic nature of the BRST cohomology
shows that the transport of \cecvafa, if it exists in
this case, must be distinct from
Lorentz transport. Moreover,
it indicates that unitarity is an important ingredient
of \cecvafa.

\subsec{Conformal perturbation theory}

The $\sigma$-model formulation of toroidal compactifications
of sec. 2.3
suggests an entirely different transport on $\IH$
in terms of conformal perturbation theory.
Naively this looks rather trivial. The operator
$$\p X \cdot \Delta E \cdot \pb X$$
is an exactly marginal operator.
The main idea of conformal perturbation theory
is that one can calculate correlators for a
theory with action $S_0+\int \Phi$ in terms
of an exponential series of correlators for a theory
with action $S_0$ involving successive
insertions of the operators $\int \Phi$.

In terms of path integrals, we may expect to
relate correlators at $E'=E+\Delta E$, where
$E',E\in\CB$,  along the following lines:
\eqn\sketchi{\eqalign{
\Biggl\langle \prod V_i(z_i)\Biggr\rangle_{E'}&\equiv
\int [DX(\sigma,\tau)] e^{-{1\over 2 \pi}\int \p X E' \pb X}
\prod V_i(z_i)\cr
&=\int [DX(\sigma,\tau)] e^{-{1\over 2 \pi}\int \p X E \pb X}
e^{-{1\over 2 \pi}\int \p X \Delta E \pb X}
\prod V_i(z_i)\cr
&=\Biggl\langle e^{-{1\over 2 \pi}\int \p X \Delta E \pb X}
\prod V_i(z_i)\Biggr\rangle_{E}\cr}
}

Of course \sketchi\ is terribly naive. For one thing,
states and operators are isomorphic in CFT, but
the statespaces $\CH_E$ and $\CH_{E'}$ are different.
Therefore, we should certainly modify the LHS of \sketchi\ to
read
$$\Biggl\langle \prod T^{E',E}V_i(z_i)\Biggr\rangle_{E'}$$
where $T^{E',E}:\CH_E\to \CH_{E'}$ is some linear transformation
of the operators at $E$ to the operators at $E'$. Note that
if \sketchi\ makes sense, then, since the path-integral
is unambiguous, the resulting parallel transport on $\CB$
must be flat.

Unfortunately,  \sketchi\ is also too naive for several
other reasons. The manipulations in \sketchi\
ignore the divergences and contact terms
which make quantum field theory nontrivial.
There are three kinds of divergences:

1. Singularities of the marginal perturbation with operator
insertions $V_i$. These are responsible for the fact that
there is a nontrivial transport $T$.

2. Singularities of the marginal perturbation with itself
in disconnected diagrams. These are responsible for a
relatively trivial renormalization of the free energy.
There are also infrared singularities, but these only
occur for the free energy.

3. Singularities of the marginal operator with itself and
with $V_i$. These contact term singularities are responsible
for the contact terms leading to different parametrizations
of moduli space.

Handling the above divergences requires introduction of a
cutoff, which typically destroys conformal invariance. It
thus might seem hopeless to use conformal perturbation theory
to define transport along a moduli space of CFT's.
We will describe two ways to get around this problem.
In this section we define the ``little-disks transport.''
In appendix A we describe a more subtle way of defining
the series.

{\bf Definition}. The little-disk cutoff configuration space
\eqn\cutcnfg{\eqalign{
F_m(z_1,\dots z_n;\epsilon)\equiv
\{(w_1,\dots w_m)\in \IC^m | i\not=j &\rightarrow |w_i-w_j|> \epsilon;
|w_i-z_j|>\epsilon,\cr
& i=1,\dots m, j=1,\dots n \}\cr}
}

{\bf Definition}: Let $\CO$ be a $(1,1)$ operator in a
CFT $\CC$. Let $\Phi_i$ be operators in
$\CC$. The point-split perturbation series is defined by
the sum of CFT correlators:
\eqn\pntsplt{\eqalign{
\Biggl\langle e^{\int_{\Sigma}
\CO}\prod_{j=1}^n\Phi_j(z_j,\bar{z}_j)\Biggr\rangle_{\CC,\epsilon}
\equiv \sum_{m\geq 0} {1\over m!} &
\int_{F_m(z_1,\dots z_n;\epsilon)} \prod_{k=1}^m d^2 w_k\qquad \qquad \cr
&\langle \prod_{k=1}^m \CO(w_k,\bar{w}_k)
\prod_{j=1}^n\Phi_j(z_j,\bar{z}_j)\rangle_{\CC}\cr}
}

In the above sum we must factor out the
disconnected terms which renormalize the free energy.
These require an infrared cutoff beyond the
ultraviolet cutoff we have specified.
We may do this by imposing the
extra condition $|z|<1/\epsilon$.
The remaining integrals are absolutely convergent.
Examples show that for appropriate $\CO$ the series
will typically be convergent.

{\bf ``Definition''}: Point-split transport:
\eqn\ptspltii{
{\langle e^{\int_{\Sigma}
\CO}\prod_{j=1}^n\Phi_j(z_j,\bar{z}_j)\rangle_{\CC,\epsilon}
\over
\langle e^{\int_{\Sigma} \CO} 1 \rangle_{\CC,\epsilon}}
\equiv \bigl\langle \prod_{j=1}^n
T^\CO_{\epsilon}\Phi_j(z_j,\bar{z}_j)\bigr\rangle
}

This definition is a little optimistic
since we have not shown that the lhs can be
expressed as a correlation function of local
operators in any theory. One may be even more
optimistic and

{\it Conjecture}: Let $\Phi_i$ be a basis of scaling
operators in the theory $\CC$. Then there is a CFT $\CC'$
(depending on $\CO$) such that
\eqn\conjlim{
T_{\epsilon}\Phi_i \sim \epsilon^{(\Delta'+\bar{\Delta'})-
(\Delta+\bar{\Delta})} T \Phi_i (1+\CO(1/\log \epsilon))
}
as $\epsilon\to 0$. Where $T:\CH_\CC\to \CH_{\CC'}$ is a
mapping of conformal field theories. The power of $\epsilon$
is required since the conformal weights change.
\conjlim\ is to be understood in the weak sense, i.e.,
as a statement about correlators, since the $\CO(1/\log \epsilon)$
corrections depend on the positions of all operators.

The conjecture is true at first order in perturbation
theory, but this is relatively trivial. Difficulties arise
already at second order in perturbation theory from
subleading logarithms. For example, there
are nontrivial contributions from regions of configuration
space where the $\epsilon$-disks overlap.
\foot{As more disks overlap the integrals for these regions
become rather tiresome.} By examining the transport of the
correlator $\langle \p Y(z_1) \p Y(z_2)\rangle$, for example, it is
easy to see that at order $\CO^2$ the strictly point
split series cannot reproduce the Lorentz-rotation
transport defined in section 7.1. The little disks
CPT is useful because it is universally and
rigorously defined, and will indeed be employed to
write a version of broken Ward identities in section 9.
Unfortunately we do not understand very well what
kind of transport it defines.

One may ask instead if it is possible to define the integrals
in CPT so that we produce the Lorentz-transport. This is
possible, and the prescription is described in detail in
appendix A.

\subsec{Zamolodchikov metric}

Suppose we have a family $\CF$ of CFT's obtained, at least
formally, by perturbing the action by $\int \lambda^i \CO_i$
where
$\CO_i$ are exactly marginal operators. These may be thought of
as tangent vectors to the  family of CFT's. In this situation
Zamolodchikov has defined a natural metric for geometry on
$\CF$:
\eqn\zamodef{ds^2_Z|_{\CC(\lambda)}
\equiv |z-w|^4 \langle e^{\int \lambda^i \CO_i}
\CO_i(z, \bar{z}) \CO_j(w,\bar{w})\rangle d \lambda^i\otimes d \lambda^j}

Indeed, considering the family of $\sigma$-models
parametrized by $E\in\CB$ we have the metric
\eqn\stanmet{\eqalign{
ds^2_Z &=
|z-w|^4 \langle \p X^\mu \pb X^\nu(z,\zb)
\p X^\rho \pb X^\xi(w,\wb)\rangle_E
d E_{\mu \nu}\otimes d E_{\rho \xi}\cr
&=tr [G^{-1} dE G^{-1} d E^t]\cr}
}
One can verify that the Zamolodchikov metric
\stanmet\ is the right-invariant metric
under the right-Mobius action of $O(D;\IR)$ on $\CB$ \rangan.

\subsec{Preferred coordinates on $\IH$}

Suppose we have a
Lorentz-rotated family which is, at least formally,
obtained by exponentiation of the exactly marginal
operator $\p Y^a \Delta\CE_{a b}\bar{\p} Y^b$.
Let us determine $\Delta\CE$ such that the Zamolodchikov
metric is the invariant metric on $\IH$, that is,
such that
\eqn\zamometii{|z-w|^4
\bigl\langle \Pi\circ T^g(\beta_{-1}^a \bar{\beta}_{-1}^{\bar{a}})(z,\zb)
\Pi\circ T^g(\beta_{-1}^b \bar{\beta}_{-1}^{\bar{b}})(w,\wb)\bigr\rangle
d(\Delta \CE_{a \bar{a}}) d(\Delta \CE_{b \bar{b}})
}
is the right-invariant Zamolodchikov metric \stanmet.
Here
$$\Pi:\CH\to \CH^{\rm exactly\ marginal}$$
is the projection from $\CH$ to the subspace of
exactly marginal operators.

\noindent
{\bf Proposition 13}: The preferred coordinate system
on $\IH$ determined by the condition that \zamometii\
coincide with the right-invariant Zamolodchikov metric is
given by:
\eqn\chgact{
\Delta\CE(g)= - 2 g_{21}^t (g_{22}^t)^{-1} \eta=
- 2\eta (g_{11})^{-1} g_{12}
}

\noindent
{\it Proof}: Use the example \expltr\ to calculate the transport
of exactly marginal operators. Next substitute \chgact\ in
\zamometii. Now
 derive a right-invariant invariant metric on $O(D)$ which
descends to the invariant metric on $\IH$ as follows.
Decompose $g\in O(D)$ as
$$g=\pmatrix{g_{11}&g_{12}\cr g_{21}&g_{22}\cr}
$$
A simple computation shows that the invariant metric on
$\IH$ is
\eqn\invmt{\eqalign{
ds^2_{inv}&=\sum_{\alpha} \tr(t^\alpha g^{-1}dg)\otimes \tr(t^\alpha
g^{-1}dg)\cr
&=-4 \tr\biggl[\eta(dg_{11} \eta g_{21}^t- dg_{12} \eta g_{22}^t)
\eta (dg_{21} \eta g_{11}^t- dg_{22} \eta g_{12}^t)\biggr]\cr}
}
where the sum is over a set of broken generators
$$t^\alpha=D \pmatrix{ 0& e_{ij}\cr -e_{ij}^{tr}& 0\cr}$$
orthogonal in the Killing metric.
Finally, use the relations of the orthogonal group
to identify \invmt\ with \zamometii.
$\spadesuit$

{\bf Remarks}

1. $\Delta \CE(g)$ and $\Delta E=E\cdot g - E$ only
agree to lowest order in perturbation theory.

2. From the formal definition \zamodef\ of the Zamolodchikov
metric we expect $\Delta \CE(g)$ to be a good set of
coordinates for conformal perturbation theory.
This is verified in appendix A.

3. Although the connection
$\nabla$ is flat, its projection to the subspace of
$(1,1)$ operators need not be flat \rangan. Indeed,
the projection is the Zamolodchikov connection, which has
nonzero curvature. This curvature can be understood to arise
from contact terms between the exactly marginal
operators
\ref\kutasov{D. Kutasov, ``Geometry on the space of
conformal field theories and contact terms,''
Phys. Lett. {\bf 220B}(1989)153}.

\subsec{Orbifold points and holonomy}

The existence of both orbifold points and a
flat connection on $\CN$ implies that the
fixed subgroups of the duality group
have holonomy representations in CFT statespaces.
In the Minkowskian case this is somewhat formal.
Nevertheless, parallel transport on $\IH$ from $E$ to
$E\cdot \gamma$ may be combined with a simple
change of basis to provide a holonomy representation
$U(\gamma)$ in $\CH_\Gamma$ where $\Gamma$ is
constructed from $E$.

Carrying out this program in the case of the
fixed subgroup $Aut(II^{1,25})\times Aut(II^{1,25})$ should be
very interesting. Perhaps a similar
procedure in the case of $\IZ_2\times \IZ_2$
orbifolds will
lead to a covariant and geometric construction of the
Fischer-Greiss Monster group.
Carrying this speculation further, we may ask
if all the sporadic groups may be so represented
as holonomy representations around orbifold points
within a single moduli space. This would provide a
unified approach to the study of finite simple groups.

\newsec{Amplitudes at Enhanced Symmetry Points}

\subsec{A philosophical digression on compactified time}

As we have stressed above, we are regarding the
finite radius of time as
an unphysical but symmetric vev of a scalar field.
Nevertheless,
the idea that time might really be periodic, albeit with an
enormous period, has been a recurring theme in human thought
in all ages and cultures
\ref\eliade{See, e.g., M. Eliade, {\it Cosmos and History:
The Myth of the Eternal Return}, Harper and Row, 1959, or
S.J. Gould, {\it Time's Arrow, Time's Cycle}, Harvard, 1987}.
We digress and comment briefly on
compactified time from a physicist's standpoint.

Classically, closed timelike loops imply that
all motion is periodic with fundamental period $T/n$
where $T$ is the length of the loop and $n$ is a positive
integer. The effects of closed timelike
loops on a physical system depend strongly on the
Hamiltonian in question. Hamiltonians which
do not have bound orbits are considered unphysical.
In general, among Hamiltonians with bound and unbound
orbits initial conditions are restricted to correspond to
some (possibly none) of the bound orbits.
Phase space is reduced to a collection of
subvarieties reflecting quantization of energy.
For example, in one-dimensional particle
mechanics with Hamiltonian $\half p^2 +V(x)$
the period integral
$$\oint {dx\over \sqrt{E-V(x)}}$$
is quantized in units of $T/n$.
The nature of the resulting quantization of $E$ depends
on the potential $V(x)$. For a square-well
potential there is a minimum energy and $E_n\sim n^2$.
For a linear confining potential there is a maximum
energy and $E_n\sim 1/n^2$. One notable exception to
these remarks is the harmonic oscillator with frequency
$\omega=n/T$. This is the unique system which imposes
no quantization at all on phase space.

It is sometimes said that closed timelike loops
imply a lack of causality. If time is globally
periodic this is not true. Closed timelike loops
merely force a further retreat from the illusion of free-will
and determinism. In classical mechanics one is only
free to choose initial conditions. Subsequent
history is completely deterministic. If time
is periodic one is no longer free to choose
initial conditions arbitrarily, but the choice might
be large and apparently arbitrary.

Quantum mechanics introduces new
problems with closed timelike loops.
To be sure, energies
are now quantized in units of $h/T$ but this
effect is in fact negligible even in the
most sensitive tests of QED. On the other hand,
serious philosophical
problems emerge connected with measurement
theory. One can, of course, simply restrict attention to
periodic solutions of the Schrodinger equation.
Nothing is measured and nothing gets done,
but life goes on. If we wish to introduce
measurements then
we can have more or less arbitrary solutions
for time period T, with periodic collapse of the
wavefunction at some specified set of times,
say, $t=n T$. In order to have true
periodicity the
result of collapsing the wavefunction must always
 be the same. This contradicts the
probabalistic interpretation of the wavefunction,
unless the $\psi$ is an eigenstate of the
observable being measured.
Something more interesting happens if we
consider two measurements at different times
of noncommuting observables. In this case
$\psi$ cannot be simultaneously an eigenstate
of both observables, the collapse of the wavefunction
must be nontrivial and the probabalistic
interpretation of $\psi$ is inconsistent
with time-periodicity. Thus, among
other things, the Stern-Gerlach experiment
shows that time  cannot be periodic.

\subsec{Genus zero string amplitudes at ESP's}

String densities defined by the operator formalism
may be thought of as differential forms on moduli space:
$\omega\in \Omega^*(\CM_{h,n};H^*(\CH_\Gamma)^{\otimes n})$
where $\Omega^*$ is the DeRham complex.
String amplitudes are, at least formally, integrals of
top-forms $\omega$ over $\CM_{h,n}$.
One naturally
asks: what are the string amplitudes for the distinguished
compactification $\Gamma_*$? The answer is that almost
{\it all} amplitudes are ill-defined.
The reason for this is that
whenever there are two symmetry states $x_1\otimes \bar{x}_1,
x_2\otimes \bar{x}_2\in H^{1,1}$ in a string amplitude
with $[x_1,x_2]\not=0$ and $[\bar x_1,\bar x_2]\not=0$,
the string density will have poles on the boundary of
moduli space where the two operators collide.
This is true because the OPE of symmetry states necessarily
contains symmetry states:
$$c \cb J^{a_1} \bar J^{\bar a_1} (1)
c \cb J^{a_2} \bar J^{\bar a_2}(2)\sim
c\p c \cb \pb \cb
f^{a_1 a_2}_{a_3} \bar f^{\bar a_1 \bar a_2}_{\bar a_3}
J^{a_3}\bar J^{\bar a_3}(2) $$
The spacetime interpretation of this result is
that states made from products of symmetry currents
have three-point interactions defined by the
structure constants of a Lie algebra. In particular,
any two on-shell states fuse to produce an
on-shell state. There is no ``violation of the
Wheeler-DeWitt constraint.'' Consequently, some
internal line must be on-shell and the amplitude
must have a pole. \foot{Special state operators in 2D string
theory are ill-defined for the same reason.}

Although the amplitudes are ill-defined, the
string {\it densities} for symmetry-states take a
remarkably simple form. In the remainder of this
subsection we digress to explain
this point, which is probably of purely mathematical
interest.

We first define some notation. The operator product
of two symmetry currents takes the form:
\eqn\crrope{
J^a(z_1) J^b(z_2)\sim \cdots +{-g^{ab}\over (z_{12})^2}+ {
f^{ab}_c J^c(z_2)\over z_{12}}+\cdots
}
The first ellipsis is in general {\it nonvanishing} because we are
working in theories where the conformal dimensions are {\it not}
bounded below. If the only operator with a one-point function
is the identity, and if the BPZ inner product is
nondegenerate, then $g^{ab}$ defines a nondegenerate invariant form
on the Lie algebra $H^1$. Let us now consider the string density
associated with a set of ``symmetry states'' $x\otimes \bar{x}$,
where $x\in H^{1,0}$ and $\bar{x}\in H^{0,1}$.
The string density is an $(n-3,n-3)$ form on $\CM_{0,n}$.
For symmetry states it factorizes holomorphically:
\eqn\strdns{
\omega(x_1\otimes \bar{x}_1,\dots, x_n\otimes \bar{x}_n)
=\tilde{\omega}(x_1,\dots, x_n)\wedge
\bar{\tilde{\omega}}(\bar{x}_1,\dots, \otimes \bar{x}_n)}
where $\tilde \omega$ is an $(n-3)$-form. Let us choose coordinates
$(z_1,\dots z_n)$ for configuration space. Fixing three points,
say, $z_1,z_2,z_3$ we define corresponding
coordinates $(z_4,\dots z_n)$ for moduli space $\CM_{0,n}$.

\noindent
{\bf Proposition 14}: A representative for the DeRham
cohomology $\tilde \omega(x_1,\dots x_n)\in H^*(\CM_{0,n})$
is given by
\eqn\reprntve{
\tilde{\omega}(x_1,\dots, x_n)=\biggl[z_{12}z_{23}z_{31}\biggr]
{1\over n}\sum_{\sigma\in S_n}
{\tr(x_{\sigma(1)}\cdots x_{\sigma(n)})\over z_{\sigma(1) \sigma(2)}
\cdots z_{\sigma(n) \sigma(1)} } dz_4\wedge \cdots\wedge dz_n
}
where $\tr$ is the nondegenerate form on $H^1$ defined by
the two-point function of currents.

\noindent
{\it Proof}:  This is easily proven by induction. First, using
the operator product expansion we deduce that the four-point
functions of currents takes the form:
\eqn\fourpt{\eqalign{
\langle J^{a_1}(z_1) J^{a_2}(z_2)J^{a_3}(z_3)J^{a_4}(z_4)\rangle
&=\qquad \qquad \qquad \cr
-\Biggl[
{f^{a_1 a_2}_b f^{b a_3 a_4}\over z_{12} z_{23} z_{34} z_{42}}
+
&{f^{a_1 a_3}_b f^{b a_2 a_4}\over z_{13} z_{32} z_{24} z_{43}}
+
{f^{a_1 a_4}_b f^{b a_2 a_3}\over z_{14} z_{42} z_{23} z_{34}}\Biggr]\cr
&+\CO\bigl({1\over z_{i4}^2}\bigr)+\CO\bigl({1\over z_{i4}^3}\bigr)+\cdots \cr}
}
By direct calculation one may check that this agrees with
\reprntve\ up to a total derivative in ${\p\over \p z_4}$.

Similarly, using the operator product expansion we may
establish the ward identity
\eqn\wrdiden{
\tilde{\omega}_n(x_1,\dots x_n)=\sum_{j=1}^{n-1}{1\over z_n-z_j}
\tilde{\omega}_{n-1}(x_1,\dots, x_{j-1},[x_j,x_n],\dots  x_{n-1})\wedge dz_n
+d \xi
}
for some $(n-4)$-form $\xi$. It is simple to check that
\reprntve\ satisfies \wrdiden\ and hence the proposition follows.
$\spadesuit$

{\bf Remarks}:

1. The sum over the permutation group in
\reprntve\ effectively antisymmetrizes the $x$'s so that the
form $\tilde\omega$ may be expressed in terms of polynomials in
the structure constants.

2. The form $\tilde\omega$ defines a map on the
tensor product of ghost number one states:
$(H_Q^1)^{\otimes n} \to H^{n-3}(\CM_{0,n})$,
or, equivalently, a map
\eqn\period{
H_{n-2}(\CM_{0,n+1})\otimes (H_Q^1)^{\otimes n+1}\to \IC
}

These maps are very interesting in view of recent developments
in the theory of BV algebras
and related structures
\ref\kontsevich{M. Kontsevich, ``Formal (Non)-Commutative
Symplectic Geometry,'' Max-Planck-Institut preprint}
\lziii\getzler\graeme
\ref\ginzburg{V. Ginzburg and M. Kapranov, ``Koszul
Duality for Operads,'' preprint}.
The top-degree homology of moduli space is:
$L_n=H_{n-2}(\CM_{0,n+1})\cong H_{n-1}(F_n(\IC))$,
where $F_n$ is the configuration space.
The collection of spaces $L_n$ together form a linear
operad known as the Lie operad:
$Lie(n)=L_n$. If $V$ is any graded vector
space then by taking invariants w.r.t. the symmetric
group $S_n$ and summing:
$$\oplus_{n\geq 0}(L_n\otimes V^{\otimes n+1})^{S_n}$$
we obtain the free Lie algebra on
$V$ \kontsevich\getzler\graeme\ginzburg.
Put another way: one can give an exotic
definition of a Lie algebra structure on $V$ as
a collection of maps
$$L_n\otimes V^{n}\to V$$
which are compatible with the operad structure.

Using the invariant
form on $H_Q^1$ (=two point function) we can write the period
maps \period\ as
$$L_n\otimes (H_Q^1)^{\otimes n}\to H_Q^1$$
Put this way, we see that the period maps can be
taken as the defining equations of the Lie algebra
$H_Q^1$. They are sure to be especially interesting
for the holomorphic CFT $\CC=S(h_{1,25}^-)\otimes \IC[II^{1,25}]$.

\subsec{Transport of amplitudes}

If we transport the symmetry states away from
a symmetric point the amplitudes become finite because
the terms in the operator product are generically off-shell,
so the BV product of \lziii\ is zero. One way to define some of the
amplitudes of the symmetry states is to rotate away from the
ESP and define the amplitude at the ESP by a limiting
procedure.

{\bf Example}: Consider vertex operators corresponding to
simple roots: $V=e^{i p_L\cdot Y}e^{i p_R\cdot Y}\CC(p_L;p_R)$
where $\CC$ is a cocycle operator. Up to a sign the four-point
function is
$$\pi
{\Gamma(p_L^{(1)}\cdot p_L^{(2)}+1)\over \Gamma(-p_R^{(1)}\cdot p_R^{(2)})}
{\Gamma(p_L^{(1)}\cdot p_L^{(3)}+1)\over \Gamma(-p_R^{(1)}\cdot p_R^{(3)})}
{\Gamma(p_L^{(1)}\cdot p_L^{(4)}+1)\over \Gamma(-p_R^{(1)}\cdot p_R^{(4)})}
$$
For simple roots
there must be either two gamma-function poles in the numerator
and one in the denominator or vice versa. Rotating away
from the symmetry point by a group element $g$ we can define
$\tilde p=g\cdot p$ and
$x_{ij}=\tilde p_L^{(i)}\cdot \tilde p_L^{(j)}-p_L^{(i)}\cdot p_L^{(j)}$.
Depending on the states at the ESP the limiting behavior
for $g\to 1$ can be any of 36 possibilities:
$x_{ij}, x_{ij}x_{ik}/x_{is}, x_{ij}/x_{is},
1,1/x_{ij}, x_{is}/x_{ij}x_{ik}$. Thus in some cases
($x_{ij}$ or $1$) the limit $g\to 1$ is well-defined.

{\bf Remarks}:

1. The string amplitudes for a fixed set of states
can be considered as functions
on $\IH$ if we use the amplitudes associated to states
all related by duality then we can construct interesting
automorphic functions on $\IH$ with prescribed
singularities at the ESP's.

2. Perhaps one could use this idea to define special state
correlators in 2D string theory.

\subsec{Higher genus amplitudes}

Any naive attempt to extend the amplitudes of
the above string theories to higher loops will
run into problems with divergences. These
divergences are unrelated to the tachyon, which,
after all, is usually {\it not} in the spectrum.
Rather, the compactification of the timelike coordinate
forces us to worry about the arbitrarily negative
conformal weights that enter in off-shell loops.
Of course, since we are working with gaussian models,
specific conformal blocks will be well-defined, but
we cannot sew them together.

In ordinary string theories this problem is
addressed in an {\it ad hoc} way.
One simply analytically
continues the gaussian integral by hand. If time
is compact one can still do this, but it essentially
amounts to an ad hoc sign change in the metric.
One loses all the interesting connections to
BRST cohomology, modular invariance etc. This is
not the right idea.

We believe the most promising direction is to give
the worldsheet a Minkowskian signature, and to
generalize G. Segal's
axiomatic framework for CFT
to the category of surfaces with Minkowski signature
conformal structure. This will require the solution
to at least three problems. First, one must decide
what singularities to allow in the metric.
Second, as we have seen in sec. 2.6,
the modular group will act ergodically on
the analog of Teichm\"uller space and one must decide
whether the theory should be formulated on moduli space
or on Teichm\"uller space.
Third,  the amplitudes will have to be understood as some kind of
generalized functions. The third point can easily be
understood for the case of the torus. Choosing the
metric to be
$$ds^2=\half \biggl[(d \sigma^1-\tau_+
d \sigma^2)\otimes (d \sigma^1-\tau_- d \sigma^2)+
(d \sigma^1-\tau_- d \sigma^2)\otimes (d \sigma^1-\tau_+ d \sigma^2)
\biggr]$$
where $(\tau_-,\tau_+)\in \IR^2 \backslash {\rm diagonal}$,
we easily calculate the path integral for a boson on radius
$R$ to be:
$$\int d \phi e^{{i\over 8 \pi}\int \sqrt{|\det g|} (\nabla \phi)^2}
={1\over |\eta|^2}\sqrt{i|\tau_--\tau_+|\over 2 R^2}\sum
e^{i \pi \tau_+ p_L^2}e^{-i \pi \tau_- p_R^2}
$$
for $\tau_->\tau_+$. The sum on momenta is a delicate
function of $\tau_\pm$. For special compactifications
and rational $\tau_\pm$ the sum can be defined by a
limiting procedure in terms of Dedekind symbols
\ref\imbimbo{C. Imbimbo, ``$SL(2,\IR)$ chern-simons theories
with rational charges and 2-dimensional  conformal field
theories,'' hep-th/9208016}.

\newsec{Broken Ward identities}

Consider a symmetry current $J$ at an
enhanced symmetry point $\Gamma$.
 Fields are organized into
representations $R$: $V^{R,\lambda}$ where
$\lambda$ is a weight vector. (The same representation
may appear with (infinite) multiplicity in the theory.)
The symmetric Ward identity is:
\eqn\symmtrc{\eqalign{
0&=\sum_i \langle \delta_J V^{R_i,\lambda_i}(z_i,\bar{z}_i)
 \prod_{j\not= i} V^{R_j,\lambda_j}\rangle\cr
\delta_J V^{R_i,\lambda_i}(z_i,\bar{z}_i)& \equiv
\oint_{z_i} dw J(w) V^{R_i,\lambda_i}(z_i,\bar{z}_i)\cr}
}

Now suppose the symmetry is spontaneously broken
by exactly marginal perturbations.
What do the Ward identities look like?

The first important point to note is that in the
broken phase
the Ward identities surely exist. In
spontaneously broken gauge theories the Ward identities,
written as Slavnov-Taylor identities, hold true
irrespective of the vacuum chosen by the theory.
Indeed this is the crucial observation behind the
proofs of renormalizability of spontaneously broken
gauge theory.

There is one very important difference in the
Ward identities in the broken and unbroken phase: in
the unbroken phase the WI's relate Green's functions
that can be taken to be simultaneously on-shell. They
are therefore identities for $S$-matrix elements.
In the broken phase this is not the case.

{\bf Example}: Consider an $SU(2)$ gauge theory.
Denote $W^\pm(p,\epsilon)$ the spacetime gauge field
of momentum $p$ and polarization $\epsilon$
associated with generators $\tau^\pm$ and $Z(p,\epsilon)$
the gauge field associated with generator $\tau^3$.
A typical Ward identity is
\eqn\typwi{\eqalign{
\langle 0|T^*\bigl[W^+(1) Z(2) W^-(3) Z(4)\bigr]|0\rangle
+&
\langle 0|T^*\bigl[W^+(1) W^-(2) Z(3) Z(4)\bigr]|0\rangle\cr
&+
\langle 0|T^*\bigl[W^+(1) W^-(2) W^-(3) W^+(4)\bigr]|0\rangle
=0\cr}
}
where $1,2,3,4$ refer to polarization and momentum.
In the unbroken phase this is true of Green's functions
for $p,\epsilon$ on-shell as well as off-shell. It is,
in particular, true of on-shell $S$-matrix elements.
In a broken phase where, typically,
$m^2_Z\not= m^2_{W^\pm}$
it is impossible to put all three terms simultaneously
on-shell.

This observation shows that it is difficult to write
the Ward identities for the broken gauge symmetries in
string theory, since we must go off-shell.
The most direct approach to this problem is to use
conformal perturbation theory with the little disks cutoff.

\noindent
{\bf Proposition 15}. (Broken Ward Identities: First version.)

Let $T_{\epsilon}=T_{\epsilon}^\CO$ the the little-disks
transport operator for a $(1,1)$ operator $\CO$ in a conformal
field theory $\CC$ defined in \ptspltii.
 If $J$ is symmetry current in $\CC$ then
we have
\eqn\brokenwi{\eqalign{
0 &= \int_{F_1(z_1,\dots z_n;\epsilon)} d^2 w \langle
T_{\epsilon}(\rho(J)\cdot \CO)(w\wb) \prod T_{\epsilon}
V^{R,\lambda}(z_i\bar{z}_i)\rangle\cr
&+\sum_i \langle T_{\epsilon}(\rho(J)\cdot V^{R,\lambda}) \prod_{j\not= i}
T_{\epsilon} (V^{R,\lambda})\rangle\cr}
}

\noindent
{\it Proof}: Expand all terms according to the definition of
the point-split perturbation series. Use the Ward identities
in the symmetric theory:
\eqn\widentiii{
0= \sum_{k=1}^m \bigl\langle \delta_J\CO_k \prod_{l\not=k}\CO_l
\prod_{j=1}^n V_j\bigr\rangle
+\sum_{i=1}^n \bigl\langle  \prod_{l}\CO_l
\delta_J V_i \prod_{j\not= i}^n V_j\bigr\rangle
}
Now use the fact that
\eqn\intgrls{
\int_{F_m(z_1,\dots z_n;\epsilon)} \prod_{i=1}^m d^2 w_i\bigl(\cdots\bigr)
=
\int_{F_1(z_1,\dots z_n;\epsilon)} d^2 w_1
\int_{F_{m-1}(z_1,\dots z_n,w_1;\epsilon)} \prod_{i=2}^m d^2 w_i
\bigl(\cdots\bigr)
}
to get \brokenwi. $\spadesuit$

The disadvantage of this approach is that the strict
little-disk cutoff gives a transport that we do not
understand very well. On the other hand, when
we interpret the Lorentz transport in conformal perturbation
theory as in appendix A, we cannot rerun the argument
of \widentiii\ because of contact terms.
Thus while it is clear from spacetime reasoning that the
broken WI's must exist they appear to be inaccessible from
the world-sheet point of view. We seem to be at an impasse.
In the next subsection we propose an end-run around the
problem.

\subsec{Transport of currents}

Consider an enhanced symmetry point $\Gamma\in \IL$ with a
holomorphic current $J$.  When the operator
$$\CO=\p Y\cdot \Delta \CE \cdot \pb Y$$
with $\Delta \CE$ given by \chgact\ is added to the action
the symmetry associated to $J$ is broken if
$$\delta_J \CO =\oint_w J(z) dz \CO(w,\wb)\not=0\qquad .$$
Let $T^g=T^{g\Gamma,\Gamma}$ be the Lorentz-transport to
some point at which $T^gJ$ is no longer an enhanced symmetry.

\noindent
{\bf Proposition 16}

(a) $T^g(\delta_J\CO)$ is a total derivative:
\eqn\trnsprtid{
T^g(\delta_J \CO)(z,\bar{z}) = - {\p \over \p \bar{z}} (T^g J(z,\bar{z})) -
{\p\over \p z}(T^g\bar J(z,\bar{z}))
}

(b) $\bar J$ may be calculated explicitly as follows.
Let $J=\CP e^{i p Y}$, where
$\CP[\beta_{-1},\beta_{-2},\dots ]$ is a polynomial in
holomorphic oscillators. Let
\eqn\defques{\eqalign{
Q_{\bar b}[\beta_{-1},\dots ]&=\sum_{s\geq 1}(-\p_z)^{s-1}
\biggl[{\p \CP\over \p \beta_{-s}^a}(z) (\Delta\CE)^a_{~\bar{b}}
e^{i p\cdot Y}(z)\biggr]_{z=w} e^{-i p\cdot Y}(w)  \cr
\tilde Q_{\bar b}[\beta_{-1},\dots ]&=\sum_{s\geq 2} (-\p_z)^{s-2}
\biggl[{\p \CP\over \p \beta_{-s}^a}(z) (\Delta\CE)^a_{~\bar{b}}
e^{i p\cdot Y}(z)\biggr]_{z=w} e^{-i p\cdot Y}(w)  \cr}
}
then
\eqn\quefrmla{
\bar{J}=Q_{\bar b}[\beta_{-1},\dots ] \bar \beta^{\bar b}_{-1} e^{i p Y}(w)
-i \tilde Q_{\bar b}[\beta_{-1},\dots ]\cdot
(\eta g_{22}^{tr}\eta g_{21} p)^{\bar b} e^{i p Y}(w)
}

\noindent
{\it Proof}: Proceed by direct calculation. Treat separately the
contractions of $\p Y(w)$ with the exponential and the
oscillator pieces:
\eqn\trnsprf{\eqalign{
\delta_J \CO(w,\wb)&=\oint_w \CP e^{i p Y}(z) dz \p Y\cdot \Delta \CE \cdot \pb
Y(w,\wb)\cr
&=\psi_1(w,\wb) + \psi_2(w,\wb)\cr
\psi_1(w,\wb)&=i \CP(w) p\cdot\Delta\CE\cdot \bar{\beta}_{-1} e^{i p\cdot Y}\cr
\psi_2(w,\wb)&=\p_w\bar{K}\cr
\p_w\bar{K}&=-:{\p \CP\over\p \beta_{-1}^a}(w) (\Delta\CE)^a_{~\bar{b}}
\bar \beta^{\bar b}_{-1}(\wb)e^{i p\cdot Y}(w):+\p_w \bar{I}\cr
\bar{I}&=\sum_{s\geq 2} \oint_w dz (-\p_z)^{s-2}\bigl({1\over z-w}\bigr)
:\biggl[{\p \CP\over \p \beta_{-s}^a}(z) (\Delta\CE)^a_{~\bar{b}} \bar
\beta^{\bar b}_{-1}(\wb)e^{i p\cdot Y}(z)\biggr]:\cr}
}
Compute $T \psi_1$ and $T \psi_2$ separately,
use the relations of $O(D;\IR)$, and take proper account of the
difference $T(\p_w\bar{J})-\p_w(T\bar{J})$ arising from normal ordering.
This gives the two terms on the right hand side of \trnsprtid\ after a
cancellation.  $\spadesuit$

\noindent
{\bf Proposition 17}: (Broken Ward identities: second version.)

\eqn\brokenwii{\eqalign{
0 &= \int_{F_1(z_1,\dots z_n;\epsilon)} d^2 w \langle
T^g(\delta_J\CO) \prod T^g V^{R,\lambda}(z_i\bar{z}_i)\rangle_{g\cdot
\Gamma}\cr
&+\sum_i \langle \biggl[\oint_{|w-z_i|=\epsilon} dw (T^g J)(w,\bar{w})
(T^g V^{R,\lambda})(z_i,\bar z_i)\biggr] \prod_{j\not= i}
(T^gV^{R,\lambda})\rangle_{g\cdot \Gamma}\cr
&+\sum_i \langle \biggl[\oint_{|w-z_i|=\epsilon} d\bar{w} (T^g
\bar{J})(w,\bar{w})
(T^g V^{R,\lambda})(z_i,\bar z_i)\biggr] \prod_{j\not= i}
(T^g V^{R,\lambda})\rangle_{g\cdot \Gamma}\cr}
}

\noindent
{\it Proof}: Use proposition 16 and integrate by parts. $\spadesuit$

{\bf Remarks}:

1. Instead of deriving identities on string amplitudes
it might be more interesting to use Proposition 17 to derive
symmetry constraints on the 1PI vertices of closed
string field theory by integrating over $\CV_{0,n}$ of
\bz.  We have not succeeded in doing
this because of the difficulties of
handling Strebel differentials.

2. Proposition 17 provides an answer to the question:
``What happens to the affine $SU(2)^{(1)}_1\times SU(2)^{(1)}_1$
symmetry of the $c=1$ gaussian model when we move away from the
self dual point?'' The answer is that to have a good
deformation theory one should look at the entire chiral
algebra. The operator product algebra of the transported
chiral currents $TJ$ closes on itself and is of index
two in the full operator algebra of the gaussian model
at finite radius. This deformation of chiral algebras is
probably related to $q$-deformed $SU(2)^{(1)}_1$, but the
details are obscure.

3. We also hope to apply the Ward identities to
high energy symmetries of string theory.
By a clever decompactification of time one might arrange
that $T^g(\delta_J \CO)$ is $Q$-trivial and $T^gJ$ is
holomorphic. If this can be done we can explain some
of the high-energy symmetries of string theory
(see the conclusions).

\newsec{Conclusions}

This paper was motivated by recent progress in CSFT and in 2D string
theory.
A comparison with the situation in two-dimensional string theory is in order.
Here too the string backgrounds leave unbroken an infinite-dimensional Lie
algebra of symmetries. It is sometimes said that
2D string theory is uniquely characterized among string backgrounds
in possessing large unbroken symmetries. Clearly, this is
not the case. It is also sometimes suggested that the vast reduction in the
number of states in the 2D case is characteristic of the existence of
large enhanced symmetries. The above examples again provide counterexamples to
this remark. The discontinuous behavior of the the BRST cohomology and the
relation to spontaneous symmetry breaking are similar in both cases.
The unbroken symmetries
in 2D are different from our case in that there is a large ghost number
zero cohomology ring (the ``ground ring''). Associated with this is
the extraordinary fact that symmetries preserve the relation $p_L=p_R$ on
tachyon vertex operators. Thus they exist at the self-dual as well as at
infinite radius, hence identities derived at the SD radius
are true at infinite radius
(at least for relations between special tachyons).
This is a much simpler situation than discussed above.

We conclude with two (wild) speculations.

First: In ordinary particle physics broken
gauge symmetries still have consequences for
physics. For example, in the standard model,
at high energies the scattering of $W$'s and $Z$'s
are related by symmetry. One may ask if there
are consequences for broken hyperbolic symmetries
when time is decompactified. The transport operator
$T^g$ maps states at fixed quantum numbers to states with
infinite energy as we decompactify.
Thus the identities in Proposition 17 should
be interpreted as {\it high energy symmetries} of string
theory. An intriguing set of speculations by D. Gross
\ref\gross{D. Gross, ``High energy symmetries of
string theory,'' Phys. Rev. Lett. {\bf 60B}(1988)1229}
and E. Witten
\ref\wittorb{E. Witten, ``Spacetime and topological
orbifolds,'' Phys. Rev. Lett. {\bf 61B}(1988)670}
seems to be closely related to these symmetries.
It would be interesting to make these remarks more
concrete. Perhaps one can use the ergodic action of
$O(\eta)\times O(\eta)$ on nearly decompactified
spacetimes.

Second:
Many facts about closed string field theory strongly indicate that the closed
string is in a spontaneously
broken phase of some more symmetric theory (e.g., topological field theory).
For example, the string field transforms inhomogeneously
$\Psi\to \Psi + Q \Lambda +\cdots$ characteristic of a broken symmetry
phase. Moreover, the action is nonpolynomial, reminiscent of the
chiral lagrangian for pion dynamics. Finally, as is clear from
the existence of enhanced symmetry points in toroidal compactifications,
Minkowski space is a very {\it unsymmetric} ground state of string.
In view of these remarks the discussion of section 5.5
suggests that the Lie algebra $\CL_U$ and its supersymmetric
analogs may be very fundamental in string theory. It further
suggests that the closed string is some kind of broken symmetry state of
a 52-dimensional open string moving in a spacetime with signature
$(\eta;-\eta)$. Recently Witten has made an interesting proposal for a
background independent formulation of open string field theory
\nref\bckind{E. Witten,
hep-th/9208027;hep-th/9210065}
\nref\liwit{K. Li and E. Witten,  hep-th/9303067}
\nref\samson{S. Shatashvili, ``Comment on the Background Independent Open
String Theory,'' hep-th/9303143}
\refs{\bckind{--}\samson}.
Rather than find an analogous formulation for closed string field theory
perhaps we should concentrate on the 52-dimensional open string and
understand the broken phase of this theory.

\bigskip
{\it Notes on the text}:

1. This paper is in final form and will not be
published anywhere else but on hep-th. Therefore, if
the reader has occasion to make reference to it,
we would appreciate it if he or she would use
the hepth number and {\it not} just the preprint number.
The same remark applies to our paper
\ref\mrsg{G. Moore, ``Gravitational Phase Transitions and
the Sine-Gordon Model,'' hep-th/9203061.}.

2. Many of our results were announced at the SUSY93
conference at Northeastern University on April 1.
There, we learned of related
independent work by K. Ranganathan, H. Sonoda, and
B. Zwiebach \rsz\  on formulating
a connection on a family of CFT's.

\bigskip
\centerline{\bf Acknowledgements}

I would like to thank several colleagues for
important and useful conversations relevant to
this material. Special thanks are due to
R. Plesser for many interesting discussions, and
for enduring many versions of appendix A below.
I also thank I. Frenkel, H. Garland, E. Getzler,
J. Horne, B. Lian, G.D. Mostow, S. Ramgoolam,
G. Segal, N. Seiberg,
G. Zuckerman, and B. Zwiebach.
This work is supported by DOE grant DE-AC02-76ER03075,
DOE grant DE-FG02-92ER25121,
and by a Presidential Young Investigator Award.

\appendix{A}{Summing the Conformal Perturbation Series}

\subsec{Preferred coordinates and conformal perturbation theory}

In this appendix we prove the following.
Let $\Gamma\in \IL$, $g\in O(D;\IR)$, and define
\eqn\chgactii{\eqalign{
\CO &=\p Y\cdot \Delta \CE \cdot \pb Y \cr
\Delta\CE(g) &\equiv -2 g_{21}^t (g_{22}^t)^{-1} \eta=
-2 \eta (g_{11})^{-1} g_{12} \cr}
}
as in Proposition 13 above. Then, for an appropriate
definition of the integrals over configuration spaces
we have:
\eqn\cptlaw{
{\langle e^{-{1\over 2 \pi}\int_{\Sigma}
\CO}\prod_{j=1}^n\Phi_j(z_j,\bar{z}_j)\rangle_\Gamma
\over
\langle e^{-{1\over 2 \pi}\int_{\Sigma} \CO} \quad 1 \rangle_\Gamma }
= \prod_{i=1}^n
\epsilon^{(\Delta_i'+\bar{\Delta}_i')-
(\Delta_i+\bar{\Delta}_i)}
\bigl\langle \prod_{j=1}^n T^g(\Phi_j)(z_j,\bar{z}_j)\bigr\rangle_{g\cdot
\Gamma}
}
where $\epsilon$ is a cutoff parameter described below and
$T^g$ is the Lorentz transport defined in section 7.
(We could absorb $\epsilon$ into $T^g$.)

Our calculation will strike many
readers as absurdly pedantic, so we offer a few
words in its defense. The essential point
is that the natural parametrization of $\IH$
follows from the contact-term prescription:
\eqn\cntpres{
\langle \p Y^a(z) \pb Y^b(\wb)\rangle = -\eta^{ab}
\pi \delta^{(2)}(z-w)
}
One might worry about multiple-contact terms when
several points coalesce. Also, \cntpres\ must be
combined with some kind of point-splitting regularization
to define the integral of $\CO(w,\bar w)$ in the neighborhood
of $\Phi_i(z_i,\bar z_i)$. We would like to specify
exactly how this should be done. Therefore,
we define the necessary integrals and the combinatorics
with some care. It may also seem silly to spend
so much effort on summing the series when we can
already calculate the answer at any point in $\IL$.
We use these known answers to learn about
how the integrals in conformal perturbation series
should be defined. The main point is:
one cannot simply integrate
over one fixed configuration space at each order; the
integrals must be defined in a more subtle way.
Roughly speaking, the treatment of the divergences on
the boundaries of moduli space depends on how we contract
operators, i.e., how the operator product expansion describes
the boundaries of moduli space.

\subsec{Proof in one-dimensional case}

We begin by considering the case of one Euclidean
signature scalar, so
we may write $\Delta\CE=-2 \tanh \lambda$ for a boost
$g(\lambda)\in O(1,1)$ as in \boost.
The corresponding operator is:
\eqn\cornrm{
\CO = -2 \tanh \lambda \p Y \pb Y }
The generalization
to several scalars is straightforward, and indicated at
the end.

Consider the numerator of \cptlaw\ and expand
the exponential series. The $m^{th}$ order term
requires a definition of the integral:
\eqn\mthterm{
\int_{F_m} \prod d^2 w_i \langle \prod_{i=1}^m \p Y\pb
Y(w_i,\wb_i)\prod_{i=1}^n \Phi_i(z_i,\zb_i)\rangle }
where $F_m$ is the configuration space of the $w_i$.

WLOG we may assume each operator is of the form
$$\Phi_i(z_i,\zb_i)=P_i[\beta_{-1},
\dots;\bar{\beta}_{-1},\dots]
e^{i p^{(i)}_L\cdot Y}(z_i) e^{i p^{(i)}_R\cdot Y}(\zb_i),$$
where
$P_i$ is a monomial formed from products of the Heisenberg modes.
Cocycle operators may be easily included and factor out of the
conformal perturbation series.

We first need to define the notion of a {\it contraction
scheme}. Consider the correlator of the $\Phi_i$. If we
evaluate it using Wick's theorem
we will use three kinds of contractions:

$A_1$. $\beta_{-n}$ in $P_i$ contracts with $\beta_{-m}$ in
$P_j$, for $i\not= j$.

$\bar{A_1}$.  Similarly for $\bar\beta$.

$B_1$.  $\beta_{-n}$ in $P_i$ contracts with $e^{i p^{(j)}_L\cdot Y}(z_j)$
for $i\not= j$. It is useful to consider the sum on $j$
of such contractions for $j\not= i$ as a single term. This eliminates
irrelevant boundary terms at infinity below.

$\bar B_1$. Similarly for antiholomorphic oscillators

$C_1$.  (Anti-)Holomorphic exponentials with (anti-)holomorphic exponentials.

Consider factoring out the result of contractions of type $C$.
The sum over the remaining terms coming from contractions of
types $A$ and $B$ is
a sum over contraction schemes. Each scheme is a combinatorial
prescription for how to contract oscillators in $\Phi_i$ with
other oscillators or with (all) the exponentials.  If there are
no exponentials the sum over contraction schemes is the same
as the sum over Wick contractions.

{\bf Example}: The holomorphic correlator $\langle \beta_{-1}e^{i p_1 Y}(z_1)
\beta_{-1}e^{i p_2 Y}(z_2) e^{i p_3 Y}(z_3) \rangle$
is a sum of {\it two} contraction schemes:
\eqn\schex{
\prod_{i<j} z_{ij}^{p_i p_j}\Biggl[{-1\over (z_{12})^2}\Biggr] +
\prod_{i<j} z_{ij}^{p_i p_j}\biggl({-i p_2\over z_{12}}-{i p_3\over
z_{13}}\biggr)\biggl({-i p_1\over z_{21}}-{i p_3\over z_{23}}\biggr)
}

Now let us return to the integrand of \mthterm.
The correlator may be
written as a sum over contraction schemes.
We now
try to relate the sum over schemes to the sum over
schemes in the original correlator.
Each term in
the sum may be visualized as a product of
chains of contractions of the operator $\CO$
beginning and ending with a contraction of $\CO$
with an oscillator or exponential in
the operators $\Phi_i$:
\ifig\chains{A chain of contractions of marginal
operators connecting two operators $\Phi_i$ and
$\Phi_j$.}{\epsfxsize 3in\epsfbox{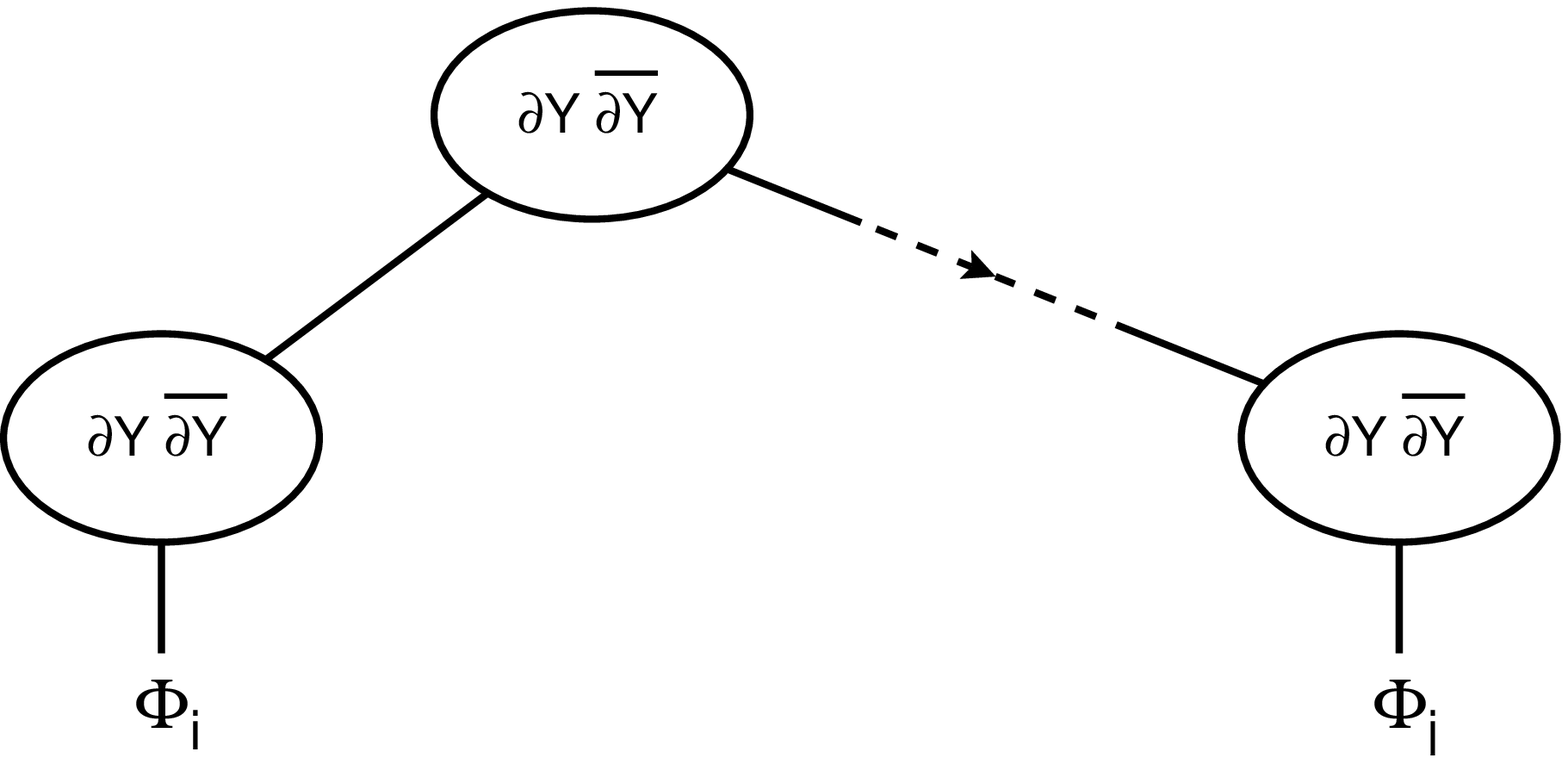}}

In each scheme the marginal operators are therefore
partitioned into disjoint chains $c_i$ which
may be regarded as
ordered subsets $\{w_{i_1},\dots w_{i_k}\}$
of $\{ w_1,\dots, w_m\}$ together with an association
of components $\beta$ or $e^{i p Y}$ from $\Phi_i$ for
initial and final points.
Thus, we may write
\eqn\sumschem{
\langle \prod_{i=1}^m \p Y\pb Y(w_i,\wb_i)\prod_{i=1}^n
\Phi_i(z_i,\zb_i)\rangle =\sum_{\CS} \prod_\ell
f_{c_\ell}(w_{i_1},\dots w_{i_{k_{\ell}}};z_1,\dots z_n)
}
where $f_{c_\ell}$ is the contribution of a chain, and we
sum over schemes.

Comparing \sumschem\ with the sum for
$\langle \prod \Phi_i\rangle$ we see that
chains in \sumschem\
may be identified with contractions in
the unperturbed correlators if the beginning and
end of the chain matches cases A-C above. We will
view the insertion of chains and subsequent integrals
over the $w$'s as a ``dressing'' of the original contraction.
In addition there are new kinds of contractions:

$A_2$: $\beta$ in $P_i$ with $\bar \beta$ in $P_j$, $i\not=j$,
and the conjugate.

$A_3$: $\beta$ in $P_i$ with another $\beta$ in $P_i$, and conjugate.

$A_4$: $\beta$ in $P_i$ with $\bar\beta$ in $P_i$, and conjugate.

$B_2$: $\beta$ in $P_i$ with the exponential at $i$ or with
an antiholomorphic exponential at $j\not= i$, and conjugate.

$C_2$: holomorphic with antiholomorphic exponentials.

$D$: Chains not involving the $\Phi$. These are cycles of
length two, corresponding to the disconnected part of
the sum.

Finally, we are ready to define the integrals in
\mthterm.
We must integrate over different spaces for different
contraction schemes. Rougly speaking, if we have a
scheme that decomposes
the set of $w_i$ into disjoint chains $c_1,\dots c_\ell$ then
we integrate over $F_{|c_i|}$ deleting  disks of radius
$\epsilon$  around
the points $z_i$ only:
\eqn\dfmthtrm{
\int_{F_m} \prod d^2 w_i \langle \prod_{i=1}^m \p Y\pb
Y(w_i,\wb_i)\prod_{i=1}^n \Phi_i(z_i,\zb_i)\rangle \equiv
\sum_\CS \prod_\ell \int_{F_{|c_\ell|}}\prod d^2 w_{i_j} f_{c_\ell}
}
We still have to define the
singular integrals over the chains. We do this with the following
integration formulae. We always assume
$\sum p_L^{(j)}=\sum p_R^{(j)}=0$. The integration
prescription is:

1.
$$\int d^2 w\Biggl(\sum_{j=1}^n {p_L^{(j)}\over w-z_j}\Biggr)
\Biggl(\sum_{j=1}^n {p_R^{(j)}\over \wb-\zb_j}\Biggr)
\equiv -\pi \sum_{1\leq j\not= k\leq n} p_L^{(j)}\cdot p_R^{(k)}
\log(|z_{jk}|^2/\epsilon^2) $$

2a. For $w_1\not= z_i$:
$$\int d^2 w\Biggl(\sum_{j=1}^n {p_L^{(j)}\over w-z_j}\Biggr)
{1\over (\wb-\wb_1)^2} \equiv - \pi
\sum_{j=1}^n {p_L^{(j)}\over \wb_1-\zb_j}$$

2b. For $n\geq 0$:
$$\int d^2 w\Biggl(\sum_{j=1}^n {p_L^{(j)}\over w-z_j}\Biggr)
{1\over (\wb-\zb_i)^{n+2}} \equiv -
\pi {1\over (n+1)!}\bigl({\p\over \p\bar{z}_i}\bigr)^n
\sum_{j\not= i}^n {p_L^{(j)}\over \zb_i-\zb_j}$$

3a. For $z_1\not= z_2$, $n\geq 0,m\geq 0$:
$$\int \prod_1^{2k+1} d^2 w_i
{1\over (z_1-w_1)^{n+2}} {1\over (\wb_1-\wb_2)^2}{1\over (w_2-w_3)^2}
\cdots {1\over (\wb_{2k+1}-\zb_2)^{m+2}}\equiv 0$$

3b. For $n,m\geq 0$
$$\eqalign{
\int \prod_1^{2k+1} d^2 w_i
{1\over (z-w_1)^{n+2}} {1\over (\wb_1-\wb_2)^2}{1\over (w_2-w_3)^2}&
\cdots {1\over (\wb_{2k+1}-\zb)^{m+2}}\cr
&\equiv\delta_{n,m}{(-1)^n\over n+1} {\pi^{2k+1}\over \epsilon^{2n+2}}\cr}
$$

4a. For $z_1\not= z_2$
$$\eqalign{
\int \prod_1^{2k} d^2 w_i
{1\over (z_1-w_1)^{n+2}} &{1\over (\wb_1-\wb_2)^2}{1\over (w_2-w_3)^2}
\cdots {1\over (w_{2k}-z_2)^{m+2}}\cr
&\equiv \pi^{2k}{1\over n+m+2} {n+m+2\choose n+1}
{1\over (z_1-z_2)^{n+m+2}}\cr}$$

4b. For $n,m\geq 0$
$$\int \prod_1^{2k} d^2 w_i
{1\over (z-w_1)^{2+n}} {1\over (\wb_1-\wb_2)^2}{1\over (w_2-w_3)^2}
\cdots {1\over (w_{2k}-z)^{2+m}}\equiv 0$$

It is important that the integrals above are related to
but differ from the corresponding point-split integrals.
The point-split integrals are easily evaluated using
$$\int_{\Sigma} d^2 w {\p\over \p \bar{w}}(f(w,\bar w))=
-{i\over 2}\int_{\p \Sigma}dw f
$$
where on the RHS one traverses the boundary with the left hand
pointing inward to the region $\Sigma$. Choosing cuts one
can evaluate the point-split integrals of type 1 exactly.
For example, for $n=2$ the exact point-split expression is
$$2 \pi p_L p_R \log\biggl( {|z_{12}|^2\over \epsilon^2}-1\biggr)$$
for $|z_{12}|>\epsilon$.
For $n>2$ integral 1 differs from the pointsplit answer by
terms of order $\CO(\epsilon^2)$.
Similarly, the other formulae are all related to the
exact point-split formula:
$$\int d^2 w {1\over ( \bar w_1-\bar w )^2}{1\over w-z}=
-\pi {1\over \bar w_1-\bar z}
$$
where we integrate over $|w-w_1|>\epsilon, |w-z|>\epsilon$ and
assume $|w_1-z|>\epsilon$. If we try to use the
point-split integrals blindly then we find the log's to
not exponentiate
and we destroy conformal invariance. We now prove that,
if we use the above definitions
of chain integrals we can preserve conformal invariance.

First note that
formlula 3a shows that chains of type $A_2$ contribute zero.
Likewise,
formula 4b shows that chains of type $A_3$ contribute zero.
The new contractions of type $A_4$ are not associated with
unperturbed contractions, but are associated with contractions
of a different unperturbed correlator. These $A_4$ contractions
are a normal ordering effect. See case 3 below.
The remaining new types of contractions: $B_2$, $C_2$
may be considered as ``dressings'' of
unperturbed contractions.

Organizing the sum this way we can examine the combinatorics.
Chains involve ordered tuples but the ordering clearly does
not affect the integrals above. Therefore
if there are $2 n_d$ chains of type D, $n_e$ chains of type
C, $n_{1},\dots n_{l}$ disjoint chains of types A,B the sum of terms
contributing to a given scheme at $m^{th}$ order in
perturbation theory will be a sum over partitions
$2n_d + n_e + \sum n_j=m$ weighted by a combinatorial factor:
\eqn\combinfctr{
{m\choose 2n_d \ n_e\ n_1\ \cdots n_l}
{(2n_d)!\over n_d! 2^{n_d}} \prod_i n_i!
}
The conformal perturbation series for the dressings of a
given contraction scheme now factors as a product of
distinct infinite series. We examine each type of series
 in turn:

1. The disconnected terms, of course, factor out as an exponential
series:
\eqn\dscseries{ exp\biggl[{\Delta \CE^2\over 8 \pi^2}\int
{d^2 w_1 d^2 w_2\over |w_1-w_2|^4} \biggr]
}
These are the only terms where an infrared as well as an
ultraviolet cutoff is necessary. The denominator in the LHS
of \cptlaw\ cancels \dscseries.

2. Contractions of type $A_1,\bar A_1$ are dressed by a
{\it geometric} series. Thus, the dressing of a contraction
of type A1 may be calculated using formula 4a:
\eqn\examplei{\eqalign{
\beta_{-n}(z_1)---\beta_{-m}(z_2)
&= \sum_{k\geq 0} \bigl({-\Delta \CE\over 2 \pi}\bigr)^{2k} \pi^{2k}
\langle \beta_{-n}(z_1)\beta_{-m}(z_2)\rangle\cr
&={1\over 1-\Delta \CE^2/4}\langle \beta_{-n}(z_1)\beta_{-m}(z_2)\rangle\cr
&=\cosh^2 \lambda\langle \beta_{-n}(z_1)\beta_{-m}(z_2)\rangle\cr}
}
where we identify $\Delta \CE=-2 \tanh \lambda$.

3. Contractions of type $A_4$ also form a geometric series
and result in new contraction schemes. These correspond
to the fact that after rotating $\beta$ in a polynomial
of $\beta, \bar \beta$'s one must normal order,
e.g., as in \expltr. For example,
consider summing the chains that connect $\beta_{-n}$
to $\bar\beta_{-n}$ at the same point $z_i$. Using formula 3b this series
becomes:
\eqn\exampleini{\eqalign{
 \beta_{-n}(z_i)---\bar\beta_{-n}(\zb_i)
&= \sum_{k\geq 0} (-{\Delta \CE\over 2 \pi})^{2k+1} n {\pi^{2k+1}\over
\epsilon^{2n}}\cr
&={-\Delta \CE/2\over 1-\Delta \CE^2/4}n {1\over \epsilon^{2n}}\cr
&=\cosh \lambda\sinh \lambda {n\over \epsilon^{2n}}\cr}
}
This corresponds to the normal-ordering in the transport:
\eqn\nrmord{\eqalign{
T^g( [\cdots \beta_{-n} \bar{\beta}_{-n} \cdots])
&= [\cdots (\cosh \lambda \beta_{-n}+\sinh \lambda \bar{\beta}_n)
(\cosh \lambda \bar\beta_{-n}+\sinh \lambda \beta_n)\cdots ]\cr
&=\cosh^2 \lambda [\cdots  \beta_{-n}\bar\beta_{-n}\cdots]\cr
&\quad +n \cosh \lambda \sinh \lambda [\cdots 1 \cdots ]\cr
&\quad + \cosh \lambda \sinh \lambda [\cdots
\bar\beta_{-n} \bar{\beta}_{n} \cdots ]\cr}
}

4. Contractions of type B. Again we get a geometric series.
Using formulae 2a,2b we see that the contraction
of $\beta_{-n}$ at $z_i$ with holomorphic exponentials
at $z_j$ is dressed as
\eqn\exampleii{\eqalign{
 \beta_{-n}(z_i)--- \prod_{j=1}^n e^{i p^{(j)} Y}
&= \sum_{k\geq 0} ({-\Delta \CE\over 2 \pi})^{2k} \pi^{2k}
\sum_{j\not= i} {-i p_L^{(j)}\over z_i-z_j}\cr
&\qquad +
\sum_{k\geq 0} (-{\Delta \CE\over 2 \pi})^{2k+1} \pi^{2k+1}\sum_{j\not= i} {-i
p_R^{(j)}\over z_i-z_j}\cr
&=\cosh \lambda \sum_{j\not= i} {-i (p_L^{(j)})'\over z_i-z_j}\cr}
}
where $p_L'=\cosh \lambda p_L + \sinh \lambda p_R$.

5. Finally we consider chains of type C. Chains of length
$k$ connect (anti-)holomorphic to (anti-)holomorphic exponentials
for $k$ even and vice versa for $k$ odd. Since we
are working with exponentials we can have an arbitrary number of
chains of length $k$ connect any two exponentials.
Suppose there are $r_i$ chains of length $2i$
connecting holomorphic to holomorphic exponentials,
$\bar{r}_i$  chains of length $2i$ connecting anti-holomorphic to
anti-holomorphic exponentials, and
$s_i$ chains of length $2i+1$ connecting
holomorphic to anti-holomorphic exponentials.
Then the conformal
perturbation series becomes the sum:
\eqn\expexp{\eqalign{
\sum_{r_i,\bar{r}_i,s_i\geq 0} \prod {1\over 2^{r_i}r_i!}
{1\over 2^{\bar{r}_i}\bar{r}_i!} {1\over s_i!}\qquad &\qquad\qquad\qquad \cr
\biggl(-{\Delta \CE\over 2 \pi}
\biggr)^{\sum_i r_i 2i +\bar{r}_i 2i +s_i(2i+1)} &
\prod\biggl(\pi^{2i+1} \sum_{i\not=j} p_L^{(i)} p_R^{(j)}\log{|z_{ij}|^2\over
\epsilon^2}\biggr)^{s_i}\cr
\prod \biggl(\pi^{2i} \sum_{i\not=j} p_L^{(i)} p_L^{(j)}\log{|z_{ij}|^2\over
\epsilon^2} \biggr)^{r_i}&
\prod
\biggl(\pi^{2i}  \sum_{i\not=j} p_R^{(i)} p_R^{(j)} \log{|z_{ij}|^2\over
\epsilon^2} \biggr)^{\bar{r}_i}= \cr
exp\Biggl[{-\Delta \CE/2\over 1-\Delta \CE^2/4} \sum_{i\not=j} p_L^{(i)}
p_R^{(j)}\log{|z_{ij}|^2\over \epsilon^2} &
+{\Delta \CE^2/8\over 1-\Delta \CE^2/4}\sum_{i\not=j} (p_L^{(i)}
p_L^{(j)}+p_R^{(i)} p_R^{(j)})\log{|z_{ij}|^2\over \epsilon^2}\Biggr]\cr}
}

Putting all these effects together, we see that we are simply
describing the sum over contraction schemes for the rotated
operators, rotated according to the Lorentz transport of sec. 7.

\subsec{Generalization to $n$-dimensional case}

Finally, if we have several Gaussian fields
$\Delta \CE $ given in \chgactii\ is the correct
parametrization with our integral prescription.
For example,
the generalization of \examplei\ with $n=m=1$ is
\eqn\exampleimat{\eqalign{
\beta_{-1}^a(z_1)---\beta_{-1}^b(z_2)
&= {-1\over (z_{12})^2}\eta\sum_{k\geq 0} ({-1\over 2 \pi})^{2k} \pi^{2k}
(\Delta \CE \eta \Delta \CE^{tr} \eta)^k\cr
&={-1\over (z_{12})^2}\eta
{1\over 1-{1\over 4}\Delta \CE \eta \Delta \CE^{tr} \eta}\cr
&={-1\over (z_{12})^2}(\eta g_{11}^{tr} \eta g_{11} \eta)^{ab}\cr}
}
where in the last line we have used the relations of $O(D;\IR)$.
Similarly, one may check that the other cases in the previous
section generalize correctly with the choice \chgactii.

\subsec{Reparametrizations}

Let us recall a well-known result, namely, that
parametrizations of a moduli
space of CFT's encodes a contact term presecription \kutasov
\ref\mss{A similar discussion appears in:
G. Moore, N. Seiberg, and M. Staudacher,
Nucl. Phys. {\bf 362} (1991) 665.}.
Suppose we consider the conformal perturbation
series for an exactly marginal operator
$\CO(z,\bar z)$. If two different ways of defining
the integral
$$\langle \biggl(\int \CO\biggr)^n \prod \Phi\rangle$$
are related by:
\eqn\twpresc{
\langle \biggl(\int \CO\biggr)^n \prod \Phi\rangle_1
=\sum_{m\leq n} C_{n,m}
\langle \biggl(\int \CO\biggr)^m\prod \Phi\rangle_2}
we may say that the two definitions differ by a
contact term prescription.
The sum $\sum_{m\leq n}$ is necessary because $\delta$-function
singularities of correlators of $\CO$ will be handled
differently in the two prescriptions. Using \twpresc\
we can express the conformal perturbation series
$$\langle e^{-\delta_1 \int \CO} \cdots \rangle_1$$
in terms of amplitudes from
definition 2. However, the resulting
series is not a new conformal perturbation series
$$\langle e^{-\delta_2 \int \CO} \cdots \rangle_2$$
unless the contact term prescription is {\it local}.
The condition to have local contact terms is that
there exists a set of numbers $D_j$ such that,
for all $n,m$
\eqn\loccond{
{1\over n!} C_{n,m}={1\over m!}\sum_{j_1+\cdots j_m=n} D_{j_1}
\cdots D_{j_m}
}
When this is satisfied $D_n={1\over n!} C_{n,1}$ and the
two conformal perturbation series are related by a
reparametrization:
\eqn\repar{
\delta_2=\delta_1 + D_2 (\delta_1)^2+ D_3 (\delta_1)^3+\cdots
}

Since the point-split series breaks conformal invariance
it cannot
be related to the definitions of this appendix by a
reparametrization of $\lambda$ or $g$. We believe that
what is going on is that the two contact term prescriptions
are not related locally, i.e., do not satisfy \loccond.

\listrefs
\bye